\documentclass[aps,10pt,pre,twocolumn,showpacs,superscriptaddress]{revtex4-1}
\usepackage{graphicx} 

\usepackage{bm,amssymb,amsmath}
\usepackage{color}
\usepackage{epstopdf}
\usepackage{xspace}
\usepackage{ulem}

\newcommand{\pe}{\text{Pe}\xspace}
\newcommand{\rhobar}{\overline{\rho}}
\newcommand{\vel}{\mathrm{v}} 

\newcommand{\citeN}[1]{\bibpunct{}{}{,}{n}{}{}\cite{#1}\bibpunct{}{}{,}{s}{}{}}

\definecolor{darkred}{rgb}{0.6,0,0}
\definecolor{darkgreen}{rgb}{0,0.5,0}
\definecolor{darkblue}{rgb}{0,0,0.5}
\definecolor{orange}{rgb}{0.8,0.4,0}

\begin{document}

\title{Freezing and phase separation of self-propelled disks}

\author{Yaouen Fily}
\affiliation{Physics Department, Syracuse University, Syracuse, NY 13244, USA}
\affiliation{Martin A. Fisher School of Physics, Brandeis University, Waltham, MA 02454, USA}
\author{Silke Henkes}
\affiliation{Physics Department, Syracuse University, Syracuse, NY 13244, USA}
\affiliation{ICSMB, University of Aberdeen, Aberdeen AB24 3UE, UK}
\author{M. Cristina Marchetti}
\affiliation{Physics Department, Syracuse University, Syracuse, NY 13244, USA}
\affiliation{Syracuse Biomaterials Institute, Syracuse University, Syracuse, NY 
13244, USA}

\begin{abstract}
We study numerically a model of non-aligning self-propelled particles 
interacting through steric repulsion, which was recently shown to exhibit active 
phase separation in two dimensions in the absence of any attractive interaction 
or breaking of the orientational symmetry.
We construct a phase diagram in terms of activity and packing fraction  and identify three distinct regimes:  a 
homogeneous liquid with anomalous cluster size distribution, a phase-separated state both at high and at low density, and a frozen phase. We 
provide a physical interpretation of the various regimes and develop scaling arguments 
for the boundaries separating them.
\end{abstract}

\maketitle


\section{Introduction}
\label{introduction}

Active materials, consisting of self-driven units capable of converting stored 
or ambient free energy into systematic movement,  provide a model for the 
dynamics of  living systems on many scales~\cite{Marchetti2013}. Examples range  
from the cytoskeleton that controls cell motion~\cite{Schaller2011}, to bacterial 
suspensions~\cite{peruani2012} and mammalian 
tissues~\cite{Poujade2007,Trepat2009}, to animal groups~\cite{Ballerini2008}. 
For the physicist, active materials are an exciting new class of nonequilibrium 
systems in which the interplay of activity, noise and interactions gives rise to 
a wealth of novel  phases with unusual structural, dynamical and mechanical 
properties. In recent years non-living analogues of active systems have been 
developed that allow controlled quantitative studies. These include 
self-propelled colloids~\cite{Palacci2010,Theurkauff2012} and 
droplets~\cite{Thutupalli2011,Sanchez2012}, as well as vibrated mechanical 
walkers~\cite{Deseigne2010}.
It was recently demonstrated that even the simplest model of an active system, 
consisting of self-propelled disks with overdamped dynamics and purely repulsive 
interactions,    exhibits surprising behavior: in a range of activity and 
density the disks phase separate into a dense fluid  phase and a gas 
phase~\cite{Fily2012,Redner2013}. The phase separation resembles that obtained 
in an equilibrium van der Waals fluid, but occurs here in the absence of any 
attractive interactions and is a direct result of the activity of the system 
that breaks detailed balance~\cite{Cates2012,Cates2013,Stenhammar2013}.  
Monodispersed active repulsive particles have also been shown to freeze  at high density into unique
 active crystals~\cite{bialke_crystallization_2012,Redner2013}.

\begin{figure}[h!]
\centering
\newcommand{\figletter}[1]{\makebox[0.49\linewidth][c]{{\small (#1)}}}
\figletter{a} \figletter{b} \\
\includegraphics[width=0.49\linewidth]{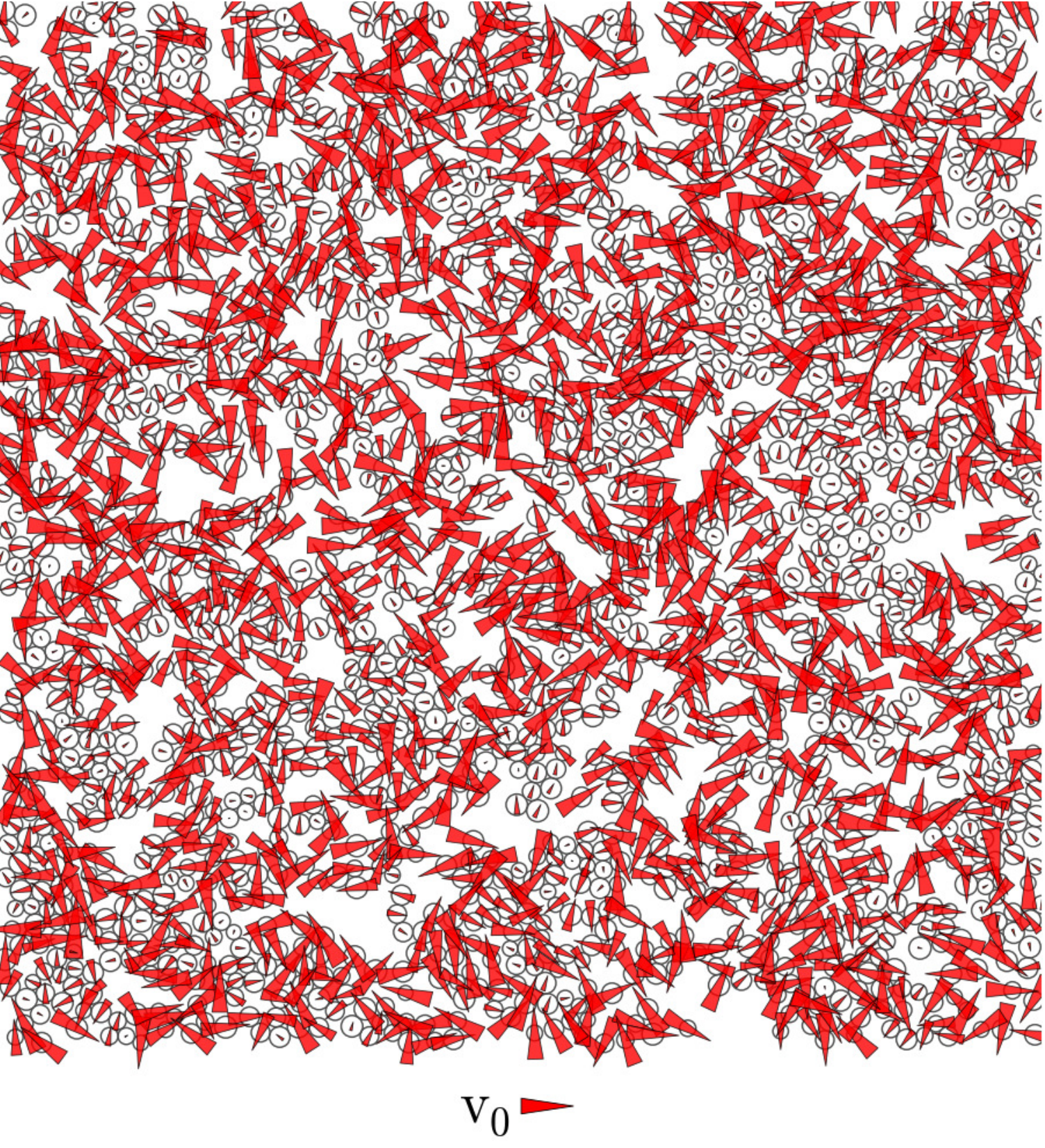}
\includegraphics[width=0.49\linewidth]{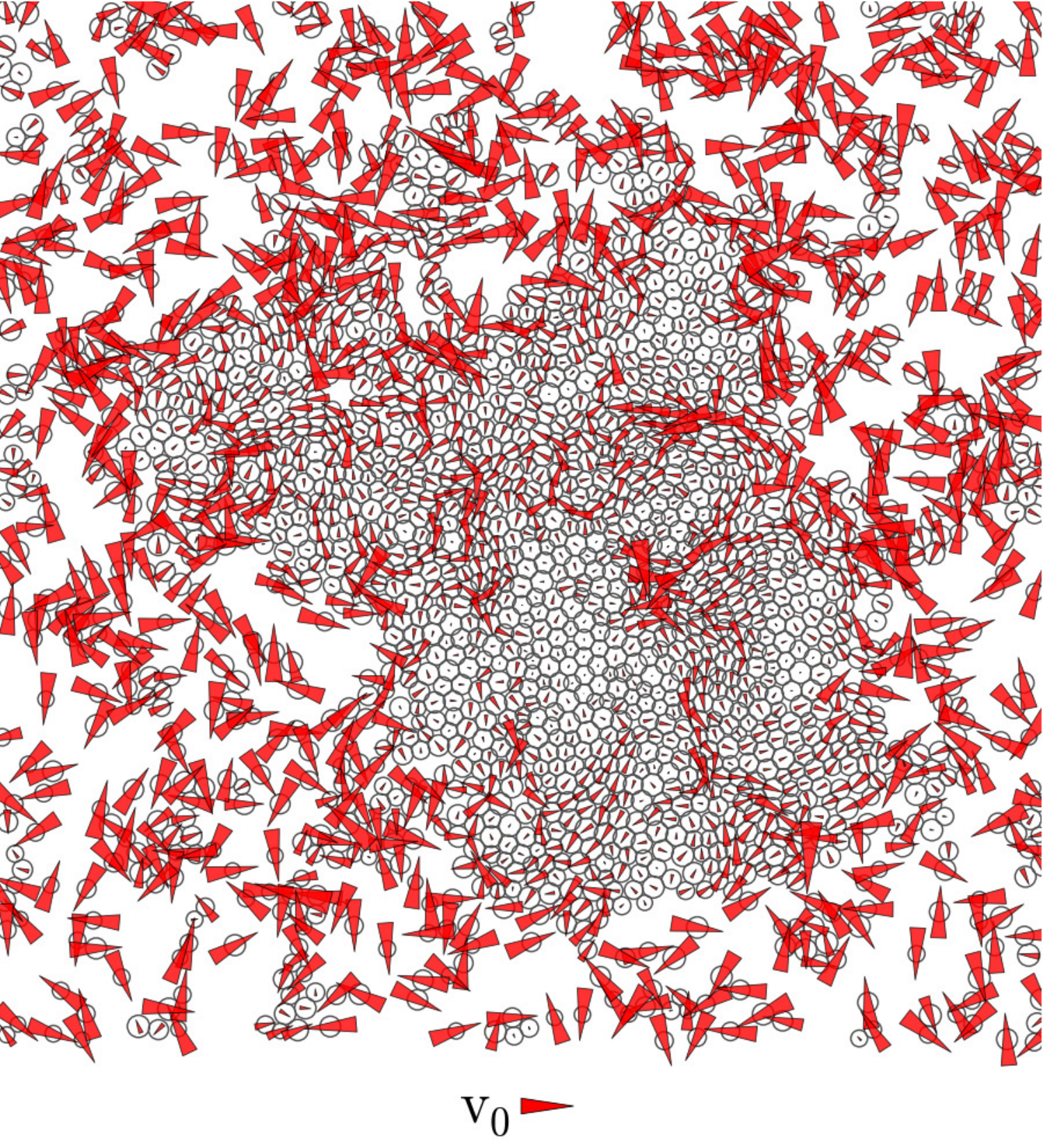} \\
\figletter{c} \figletter{d} \\
\includegraphics[width=0.49\linewidth]{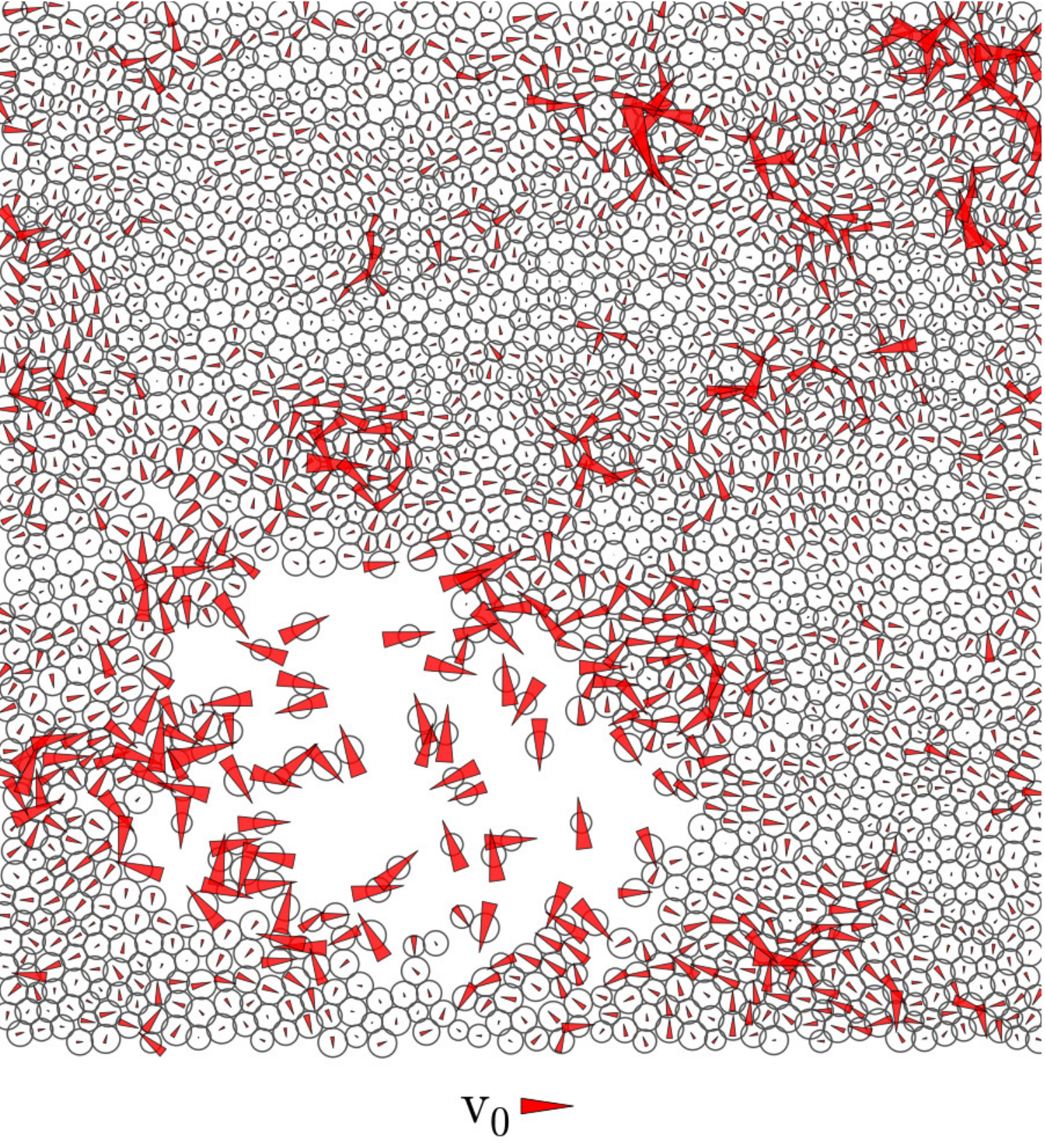}
\includegraphics[width=0.49\linewidth]{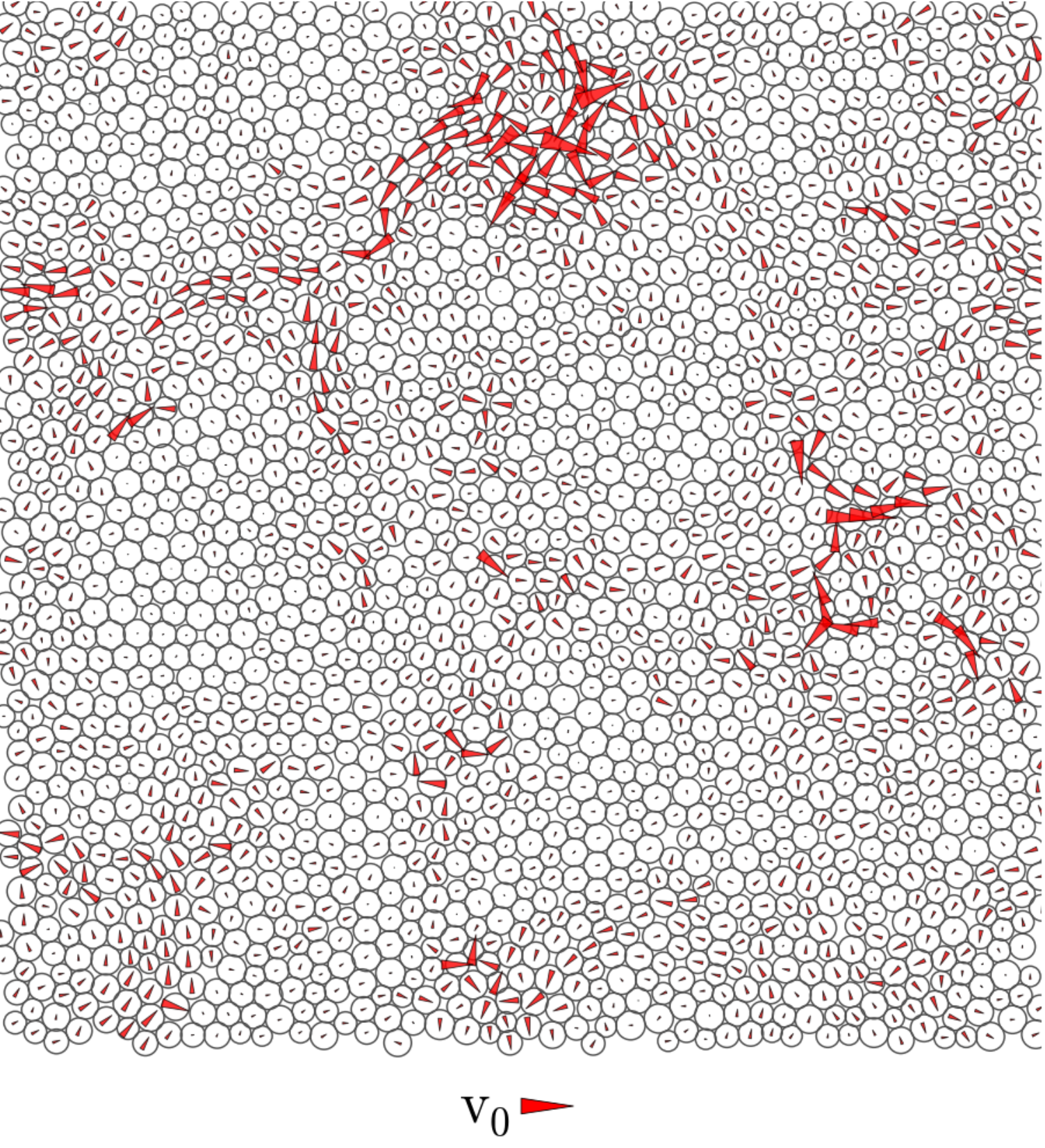} \\
\figletter{e} \\
\includegraphics[width=0.49\linewidth]{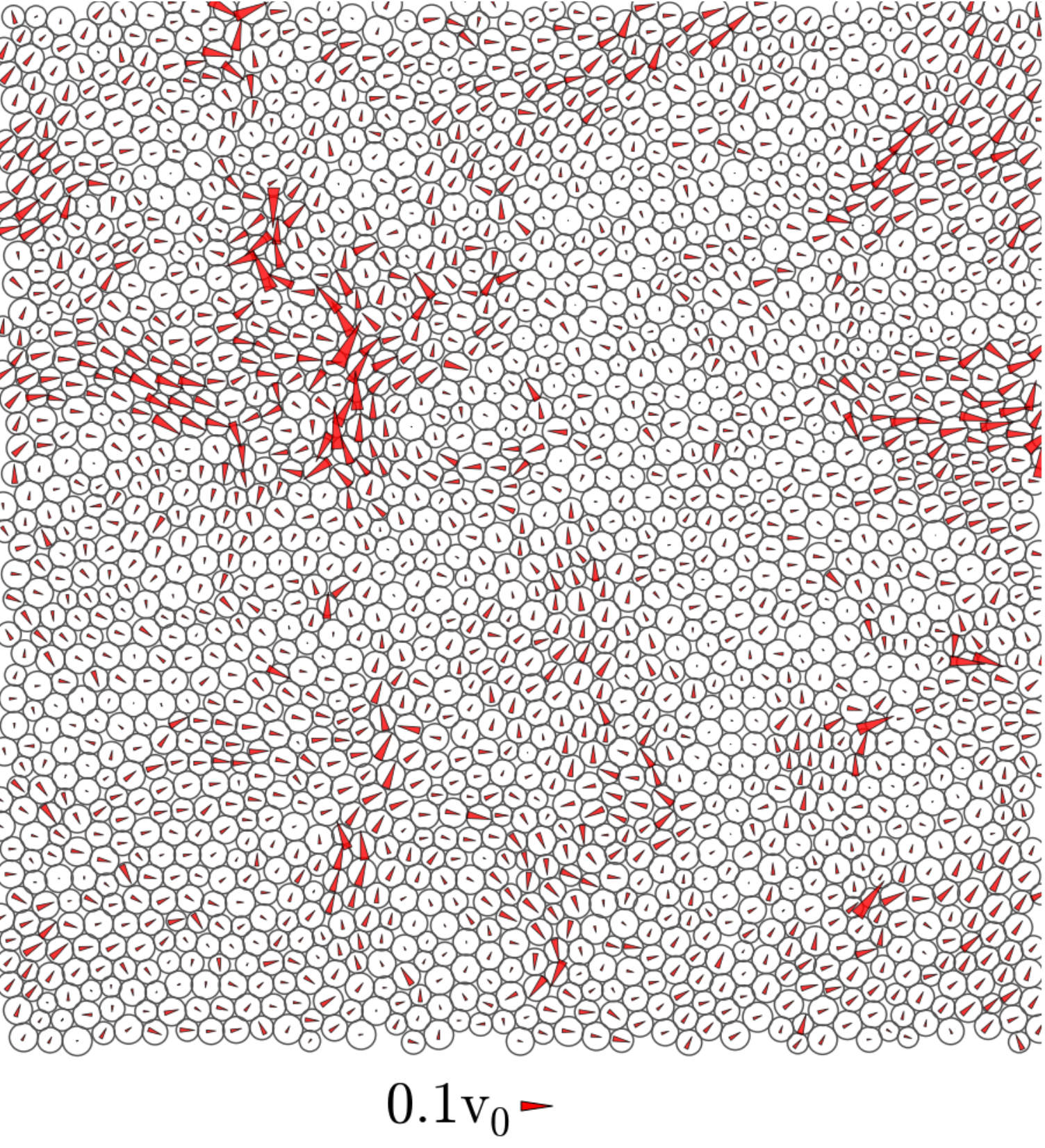}
\caption{
Snapshots of the system in four regions of phase space, for 
$\tilde{\nu}_r=5.10^{-4}$.
(a) homogeneous fluid below close packing at low self-propulsion speed ($\tilde{\vel}=0.001$, $\phi=0.5$);
(b) phase separated state below close packing ($\tilde{\vel}=0.1$, $\phi=0.5$), consisting of a high density liquid cluster surrounded by a gas of active particles;
(c) phase separated state above close packing ($\tilde{\vel}=0.1$, $\phi=0.9$), consisting a high density liquid surrounding a hole filled by a gas of active particles;
(d) homogeneous liquid above close packing ($\tilde{\vel}=0.025$, $\phi=0.9$);
(e) glassy phase ($\tilde{\vel}=0.005$, $\phi=0.9$).
Velocities are shown as red arrows, whose scale is indicated below each plot.
See also the movies 1(a) through 1(e), corresponding to each snapshot, in the Supplementary Material.
}
\label{fig:snaps}
\end{figure}

In this paper, we explore systematically the phase diagram of purely repulsive, polydisperse active particles
in two dimensions by covering a wide range of parameters in the regime where 
non-equilibrium effects are expected. We track the onset of freezing and phase 
separation by monitoring the mean square displacement and number fluctuations, 
supported by visual observations of the state of the system (see typical 
snapshots in Fig.~\ref{fig:snaps}).

We identify three regimes in the phase diagram of the system, as shown in 
Fig.~\ref{fig:PD_cmap}: a  homogeneous fluid, a phase separated fluid, and a 
glassy phase. Using scaling arguments combined with the mean-field model 
developed in a previous paper~\cite{Fily2012}, we  provide a phenomenological 
estimate for the low density boundary for the onset of phase separation which is in qualitative 
agreement with the data. As argued in the literature~\cite{Cates2012,Fily2012, Redner2013,Cates2013}, this boundary may be interpreted as a spinodal line.
This view is supported by the analysis of the cluster size distribution (CSD), which shows 
a diverging typical cluster size as the transition is approached.
Non-thermal features in the CSD persist away from the transition and present similarities with what has been seen in other active systems~\cite{Peruani2006,Peruani2013}.
Drawing on the liquid-gas analogy to equilibrium systems, 
the lowest activity strengths at which phase separation is observed may be 
identified as a critical point in the density-activity phase diagram, although more work is needed to firmly establish this. 

The phase separated region, previously identified below close 
packing~\cite{Fily2012, Redner2013}, is found to persist above close packing. In 
this case, in a reverse-clustering phenomenon,  a low-density hole spontaneously 
forms while active forces push the rest of the system into an over-compressed 
state.
This ``hole'' phase is separated from the glassy phase by a homogeneous liquid 
phase (see Fig.~\ref{fig:PD_cmap}).
As activity is increased at constant density above close packing, the system 
first goes through a melting transition from a frozen solid into a homogeneous 
fluid before reaching a distinct second transition to a phase separated state 
(see Fig.~\ref{fig:v-MSDexp-NFexp}).

At very high activity, a weakly interacting limit is reached and clustering 
disappears. This crossover, however, is not generic as it depends on the type of 
repulsive interaction.

Section 2 introduces our model of self propelled soft disks and provides a
detailed discussion of the connection to thermal systems and to  athermal
soft sphere packings in the zero-activity limit. 
In Section 3 we identify various phases in terms of the behavior of the mean square displacement and of number fluctuations and describe the phase diagram of the model. 
We also discuss the behavior of the cluster size distribution near phase separation.
Section 4 develops a mean-field theory of phase separation in the region below close packing,
including predictions for the low-density portion of the spinodal line. 
We end the section with a discussion of the freezing and high-density phase separation transitions and present scaling arguments for the corresponding phase boundaries.

\section{Model}
\label{model}

We consider a two-dimensional system of $N$  colloidal particles in an area 
$L\times L$,  modeled as disks~\cite{Fily2012}. The dynamics of the $i$-th disk is 
described by the position ${\bf r}_i$ of its center and the orientation 
$\theta_i$ of a polar axis $\hat{\bm n}_i=(\cos\theta_i,\sin\theta_i)$. The 
dynamics is overdamped and is governed by the equations
\begin{subequations}
\label{eq:model}
\begin{gather}
\label{ri}
\partial_t{\bf r}_i=\vel_0\hat{\bm n}_i+\mu\sum_{j\neq i}{\bf F}_{ij}\;,\\
\label{thetai}
\partial_t\theta_i=\eta_i(t)\;,
\end{gather}
\end{subequations}
with $\vel_0$   the  single-particle self-propulsion speed and $\mu$ the mobility. 
The angular dynamics is controlled entirely by the Gaussian white rotational 
noise $\eta_i(t)$ with zero mean and correlations 
$\langle\eta_i(t)\eta_j(t')\rangle=2\nu_r\delta_{ij}\delta(t-t')$, where $\nu_r$ 
is the rotational diffusion rate. For simplicity, we neglect translational noise 
in Eq.~\eqref{ri}, although its effect is described in the mean-field model below
(see section~\ref{mean-field}).
The $i^\text{th}$ disk has radius $a_i$ and the radii are uniformly distributed 
with mean $a$ and $20\%$ polydispersity.
The particles interact through soft repulsive forces $\mathbf{F}_{ij}=F_{ij}\mathbf{\hat{r}}_{ij}$, with 
$\mathbf{\hat{r}}_{ij}=(\mathbf{r}_i-\mathbf{r}_j)/r_{ij}$, $r_{ij}=|\mathbf{r}_i-\mathbf{r}_j|$, and ${ 
F}_{ij}=k(a_i+a_j-r_{ij})$ if $r_{ij}<a_i+a_j$ and ${F}_{ij}=0$ otherwise.
The interactions are radially symmetric and do not directly couple to the 
angular dynamics. 
The system is simulated in a square box with periodic boundary conditions, and 
the box size $L$ is adjusted to obtain the desired packing fraction $\phi=\sum_i 
\pi a_i^2/L^2$.

The model contains three time scales: $a/\vel_0$ is the time it takes a 
free particle to travel its own radius, $\tau_r\equiv\nu_r^{-1}$ is the 
correlation time of the orientation, and $(\mu k)^{-1}$ is the elastic time 
scale.
We render the equations of motion dimensionless by using $a$ and 
$(\mu k)^{-1}$ as units of length and time, respectively.
The parameter space is then explored by varying  the scaled single-particle 
self-propulsion speed $\tilde{\vel}\equiv \vel_0/(a\mu k)$, which controls the typical 
overlap due to activity, the scaled rotational diffusion constant 
$\tilde{\nu}_r\equiv \nu_r/(\mu k)$, and the packing fraction $\phi$.
A further important dimensionless quantity is the P\'eclet number 
$\pe=\vel_0/(a\nu_r)=\tilde{\vel}/\tilde{\nu}_r$, which describes the distance 
travelled by a free particle before it loses its orientation.
There are three limiting cases in which the system exhibits familiar behavior.

First, when $\tilde{\vel}\gg1$ interactions are negligible compared to 
active forces. Our soft disks can then pass through each other, behaving as a 
collection of free self-propelled particles, each performing an independent 
persistent random walk.

Secondly, in the absence of activity ($\tilde{\vel}\rightarrow 0$) the 
system reduces to an athermal packing of soft spheres. This system is well known 
to exhibit a kinetic arrest with increasing density \cite{O'Hern2003,Olsson2007,Ikeda2012}, although the onset of 
rigidity depends on the protocol used to generate the dense packing. Numerical 
studies of the rheology of packed spheres have shown that the onset of a glassy 
state at a packing fraction $\phi_G$ upon cooling is distinct from the jamming 
obtained in the limit of zero strain rate at 
$\phi_{RCP}>\phi_G$~\cite{Ikeda2012}. For polydisperse repulsive disks 
$\phi_G\simeq 0.8$ and $\phi_{RCP}\simeq 0.84$. Recent numerical work on active 
repulsive particles in both two and three dimensions  has shown that activity 
shifts the glass transition to higher packing fractions, allowing the study of 
packings close to $\phi_{RCP}$~\cite{Henkes2011,Berthier2013b,Ni2013}.

Finally, when $\nu_r\!\rightarrow\!\infty$ activity 
becomes equivalent to a white Gaussian translational noise~\cite{Fily2012}.
The system then maps to an equilibrium thermal fluid with an effective 
temperature $T_{a}=\vel_0^2/(2\nu_r)$.
This limit is only realized, however, if the orientational correlation time
$\tau_r=\nu_r^{-1}$  is much smaller than all time scales present in the system.
In particular, $\tau_r$ must be small compared to the mean free time between collisions,  $\tau_c\simeq(2a\vel_0\rho)^{-1}$, with $\rho=N/L^2$ the areal density.
The thermal limit is obtained for $\zeta=\tau_r/\tau_c\simeq 2\pe\phi/\pi\ll1$. We did not, however, simulate this range of parameters which is prohibitively time consuming. The onset of phase separation has also been interpreted in term of a critical value of $\zeta$~\cite{Redner2013}, and
a recent study pointed out the crucial role played by $\zeta$ in active suspensions, where hydrodynamic interactions are important~\cite{Fielding2013}.
Finally, the thermal limit also requires $\tau_r$ to be much smaller than the elastic time scale $\tau_e=(\mu k)^{-1}$, i.e., $\tilde{\nu}_r=\tau_e/\tau_r\gg1$. 

In the following we mainly explore parameter values in the regime $\zeta>1$ where the non-equilibrium dynamics responsible for phase separation is expected to control the behavior.
Typical runs simulate $N=2000$ particles during a time $t=10^4$ in scaled units, 
at densities $0.1\le\phi\le1.2$, scaled velocities $10^{-3}\le\tilde{\vel}\le10$ 
and rotational diffusion $0\le\tilde{\nu}_r\le10^{-2}$. 
The quantities of interest are averaged over the second half of the run as well 
as over two independent configurations, i.e. two different realizations of the 
random initial positions, angles and radii of the particles, and angular noise.
For the runs performed in the absence of angular noise ($\tilde{\nu}_r=0$), 
we lose ergodicity since time averaging does not average over particle 
orientations which are frozen for the whole duration of the run. This is 
addressed by using a higher number of independent configurations ($10$ instead 
of $2$) for these runs. Ergodicity is also lost for glassy configurations, 
however as long as $\tilde{\nu}_r$ is finite, the system still explores the 
local cage structure around each particle
\footnote{We have excluded runs at $\tilde{\nu}_r=0$ in the glassy 
region from the analysis}.

\section{Phase separation and melting}
\label{MSD}
\label{NF}

\begin{figure}[h]
\centering
\includegraphics[width=0.99\linewidth]{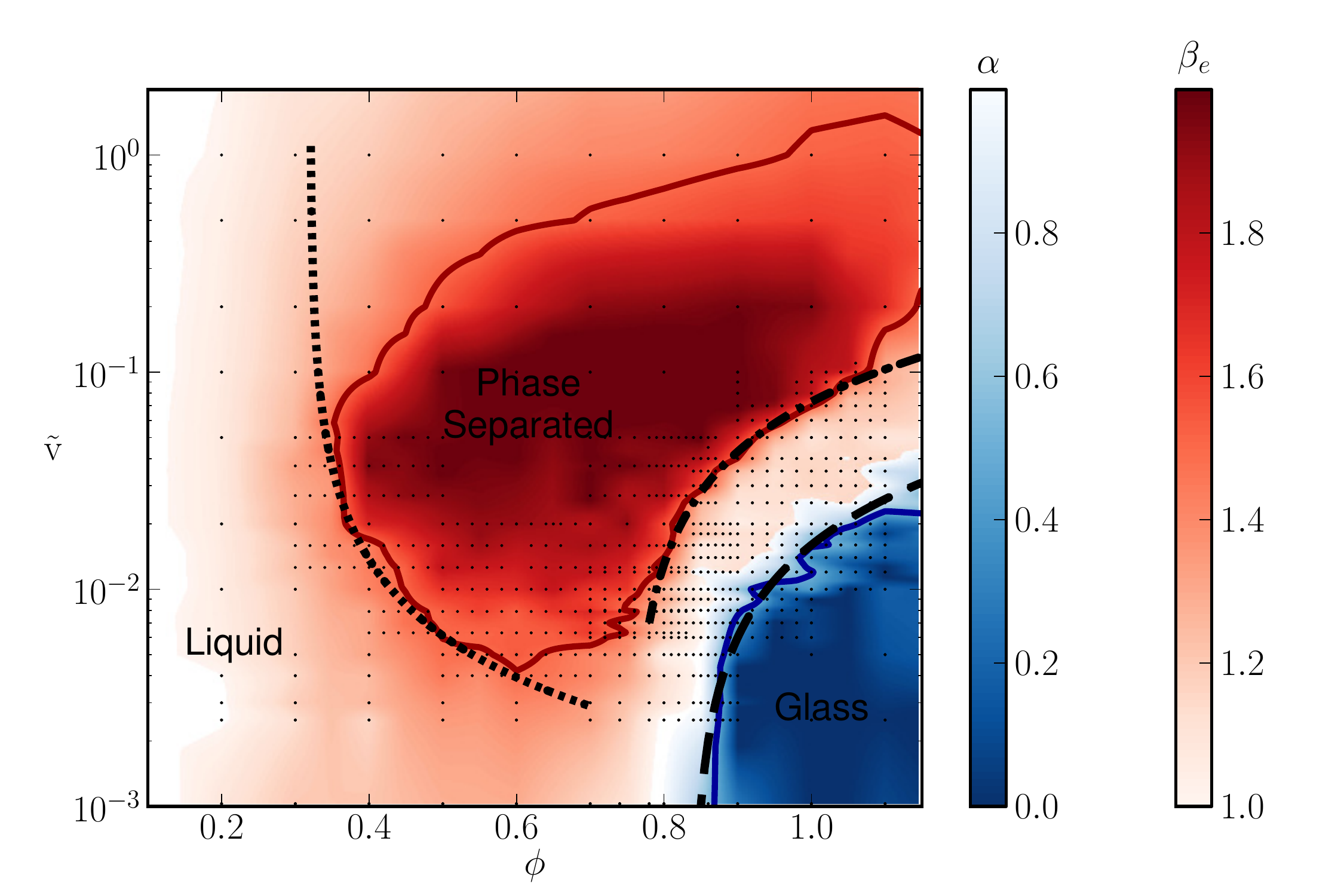}
\caption{
Color map of the exponents $\alpha$ (MSD) and $\beta_e$ (number fluctuations) in the 
$(\phi,\tilde{\vel})$ plane for $\tilde{\nu}_r=5\times10^{-4}$ showing the three 
phases of the system: homogeneous liquid, phase separated, and glassy. The 
boundary of the glassy (resp. phase separated) phase is defined as the set of 
points where $\alpha=0.5$ (resp. $\beta_e=1.5$).
The dotted line is the mean-field spinodal line given by Eq.~\eqref{eq:phis} for $D=0$ and
corresponds to $\phi=\phi_1+\phi_2/\pe$ where \pe is the P\'eclet number.
$\phi_1=0.3$ and $\phi_2=2.2$ are fitted to match the lower left side of the 
separated phase boundary (see section~\ref{mean-field}).
The dashed and dotted-dashed lines are linear fits to the melting line and lower 
right side of the separated phase boundary, respectively (see 
section~\ref{above}).
}
\label{fig:PD_cmap}
\end{figure}

The phase behavior  of the system, as obtained from simulations of our model, is 
summarized in Fig.~\ref{fig:PD_cmap}. We identify three regimes: a homogeneous 
fluid phase, a frozen state at high packing faction and relatively low activity, 
and a regime where the system is phase separated. The phase separated regime exists in an intermediate range 
of packing fraction and activity and it changes continuously from a high density cluster 
surrounded by a gas at packing fractions below close-packing, as shown in Fig.~\ref{fig:snaps}b, 
to a ``hole'' of gas phase inside a densely packed liquid at packing fractions above close-packing, as shown in Fig.~\ref{fig:snaps}c.

\subsection{Frozen state and active melting}

\begin{figure}[h]
\centering
\includegraphics[width=0.9\linewidth]{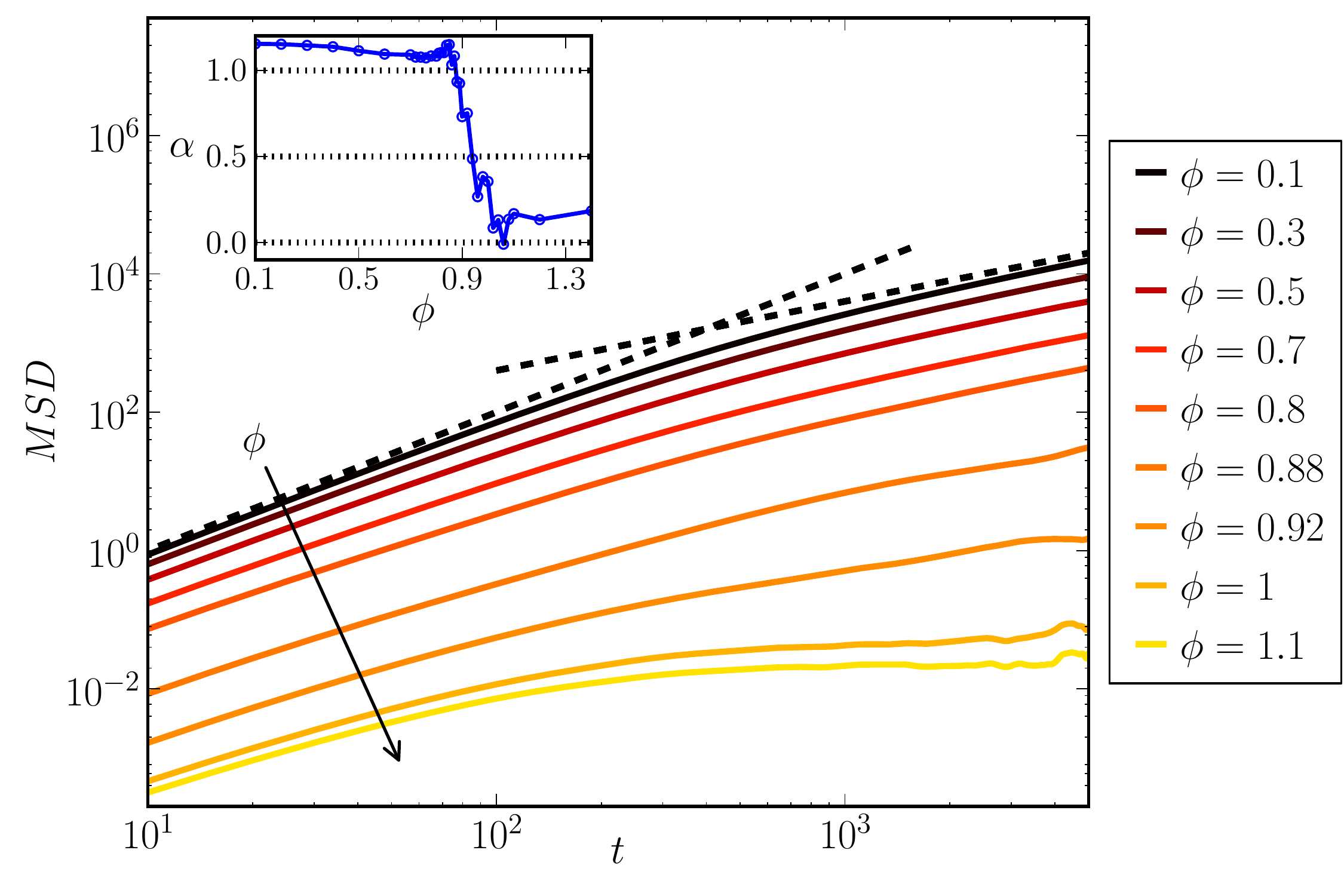}
\caption{
Mean square displacement (MSD) as a function of time for 
$\tilde{\nu}_r=5\times10^{-4}$, $\tilde{\vel}=10^{-2}$, and various values of 
$\phi$ showing both the ballistic short-time behavior and the long-term 
diffusive (for small $\phi$) or bounded (for large $\phi$) behavior. The dashed 
lines corresponds to slopes $1$ (diffusive) and $2$ (ballistic) respectively.
Inset: Exponent $\alpha$ obtained by fitting the long-time MSD to a power law.
}
\label{fig:MSD}
\end{figure}

To quantify the onset of the frozen state we have evaluated numerically the mean 
square displacement (MSD) $\langle\left[\Delta r(t)\right]^2\rangle=N^{-1}\sum_i 
\left[{\bf r}_i(t)-{\bf r}_i(0)\right]^2$ in the center of mass 
frame~\footnote{Since momentum is not conserved, the center of mass is not 
fixed. It moves with velocity ${\bf v}_{cm}=\vel_0\sum_i {\bm n}_i$ and $v_{cm}\sim 
\vel_0 N^{1/2}$.
In a finite-size frozen or almost frozen state, this uniform drifting motion 
becomes the dominant contribution to the MSD and needs to be subtracted 
out.}.
An individual self-propelled disk performs a persistent random walk, with MSD 
given by 
\begin{equation}
\label{eq:MSD}
\langle\left[\Delta r(t)\right]^2\rangle=4D_0\left[t+\frac{e^{-\nu_r 
t}-1}{\nu_r}\right]\;,
\end{equation}
with $D_0=\frac{\vel_0^2}{2\nu_r}$ the diffusion coefficient. The MSD is ballistic 
at short times and diffusive for $t\gg\nu_r^{-1}$, where $\langle\left[\Delta 
r(t)\right]^2\rangle\sim 4D_0t$. It was shown in Ref.~\citeN{Fily2012} that at 
intermediate packing fractions, before the system phase separates, the MSD can 
still be fitted by the form given in Eq.~\eqref{fig:MSD}, but with a 
renormalized local self-propulsion speed $\vel_0\rightarrow \vel(\rho)$ that accounts 
for the motility suppression due to caging by neighboring disks and an effective 
diffusivity $D_e(\rho)=\frac{\vel(\rho)^2}{2\nu_r}$. This mean-field type 
approximation works well provided the self-propelled particles experience many 
collisions before their direction of persistent motion is randomized by 
rotational diffusion, i.e., in the regime $\zeta>>1$. 

The MSD evaluated numerically over a range of packing fractions is shown in 
Fig.~\ref{fig:MSD} as a function of time. Although the effective diffusion 
constant is greatly suppressed at intermediate packing fraction, the long-time 
behavior remains diffusive. 
At high packing fraction and low activity, however, the MSD is bounded, 
indicating a frozen state. The boundary between fluid and frozen state is 
identified by measuring the  exponent $\alpha$ controlling the long-time 
behavior of the MSD, $\langle\left[\Delta r(t)\right]^2\rangle\sim t^\alpha$ and 
choosing a threshold value $\alpha_x=0.5$ to separate the ``frozen'' state 
($\alpha<\alpha_x$) from the liquid state ($\alpha>\alpha_x$). We note that an 
identification of the transition based on a drop of the long time diffusion 
constant or simply the final value of the MSD would yield similar results. 
The results are displayed in the phase diagram of Fig.~\ref{fig:PD_cmap} 
in the form of a color plot of the exponent $\alpha$ in the $(\phi,\tilde{\vel})$ 
plane. The blue solid line is the locus of points where $\alpha=\alpha_x$.

\subsection{Phase separation}

\begin{figure}[h]
\centering
\includegraphics[width=0.9\linewidth]{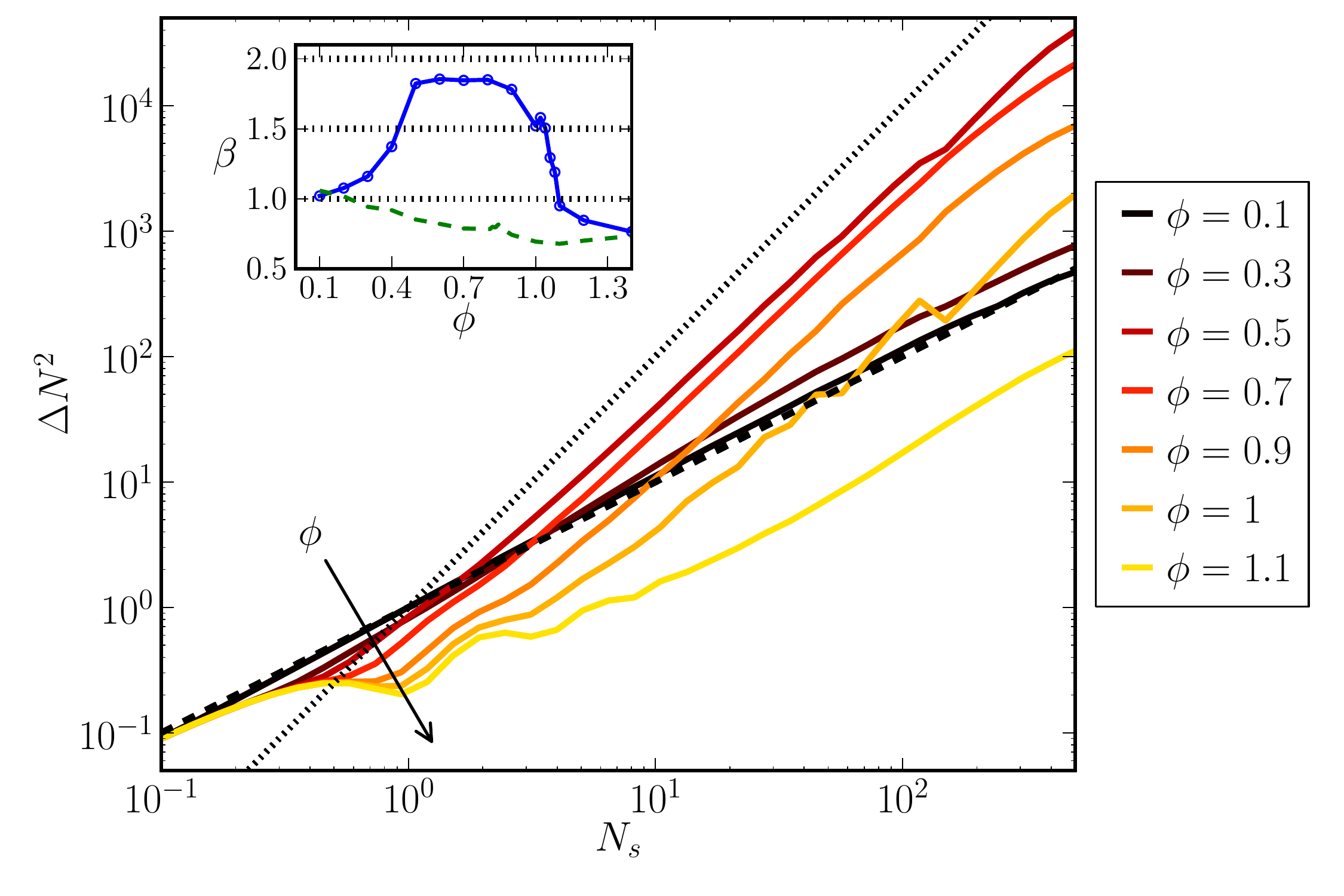}
\caption{(color online)
Number fluctuations $\left(\Delta N\right)^2$ for 
$\tilde{\nu}_r=5\times10^{-4}$, $\tilde{\vel}=10^{-1}$, and various values of 
$\phi$ showing a linear behavior at low density, close to quadratic in the phase 
separated region and sublinear in the glassy phase. The dashed and dotted lines 
correspond to slopes $1$ (ideal gas) and $2$ (phase separated), respectively.
Inset: The solid line (blue online) shows the value of $\beta$ obtained 
by the power-law fit $\Delta N^2(N_s)\sim N_s^\beta$ of the data for 
$\tilde{\vel}=0.1$  at large $N_s$. The dashed line  (green online) shows the 
corresponding exponent $\beta_0$ obtained in thermal systems ($\tilde{\vel}\rightarrow 0$) 
where self-propulsion is replaced by translational thermal noise at a temperature $T$, 
in the limit $\tilde{T}\rightarrow 0$, where $\tilde{T}=k_B T/(ka^2)$. 
The result displayed is the average of runs for $\tilde{T}=10^{-6}$, $2\times10^{-6}$ and $5\times10^{-6}$.
The difference between the two exponents displays the  hyperuniform nature of 
the fluctuations in the packed state. 
}
\label{fig:NF}
\end{figure}

\begin{figure}[h]
\centering
\includegraphics[width=0.99\linewidth]{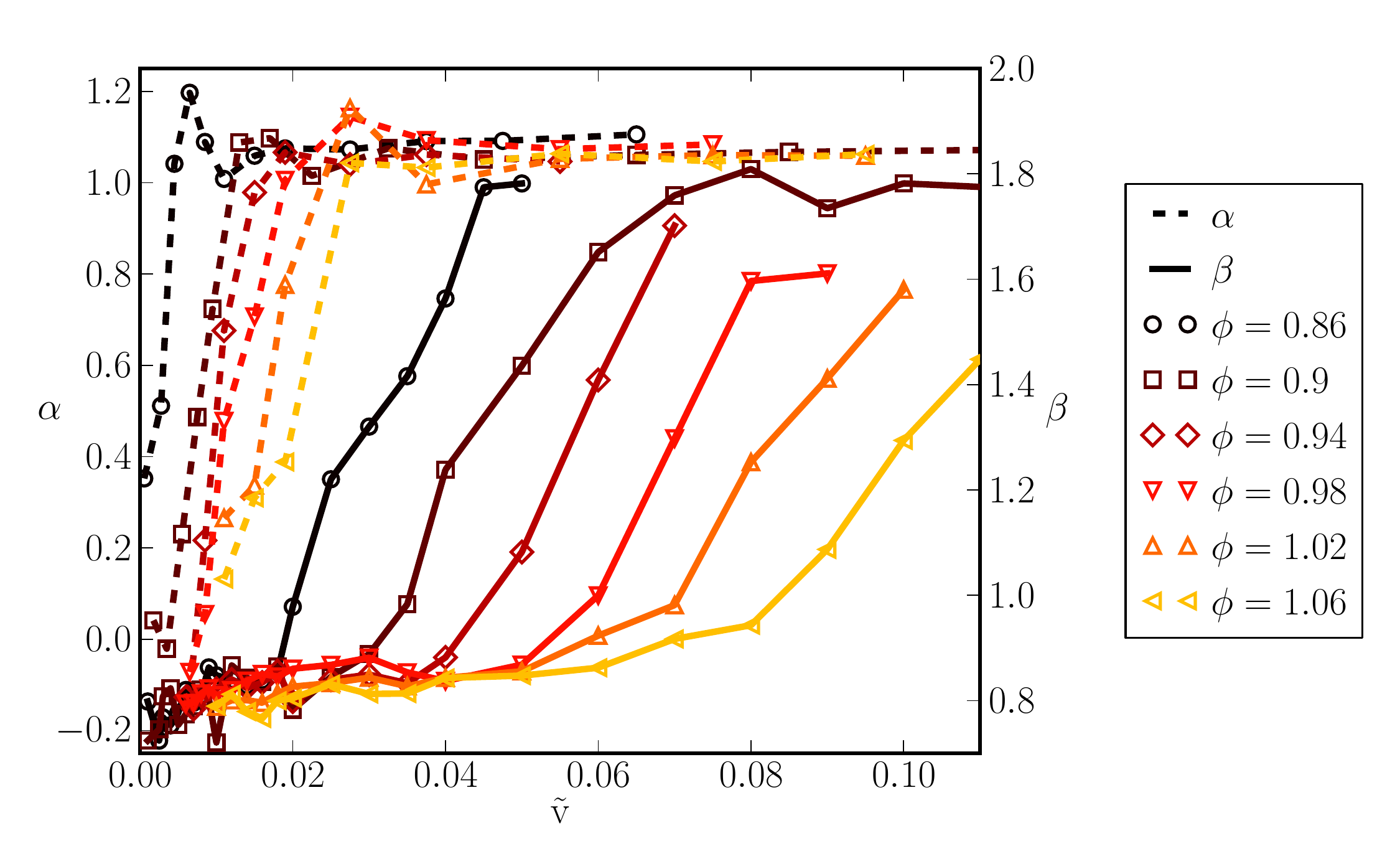}
\caption{
The long time MSD exponent $\alpha$ (dashed lines) and the number fluctuation 
exponent $\beta$ (solid lines) as functions of the self-propulsion speed $\tilde{\vel}$ 
for a range of packing fractions $\phi$ above random close packing. 
Note the convergence to $\beta \approx 0.8$ in the low $\tilde{\vel}$ limit due to hyperuniformity. 
The jump in the exponents signal the melting transition 
($\alpha$) and the the onset of phase separation ($\beta_e$).
The figure shows that for each fixed value 
of $\phi$ melting and phase separation occur at different values of $\tilde{\vel}$.
}
\label{fig:v-MSDexp-NFexp}
\end{figure}

To identify the phase separated region, 
we have measured  the number fluctuations, i.e. the spatial variance 
$\langle\left[\Delta N\right]^2\rangle $ of the number of particles in a 
subsystem as a function of the average number $N_s$ of particles in the 
subsystem. The results are shown in Fig.~\ref{fig:NF} for a range of packing 
fractions. 
For large subsystem sizes, a power law $\langle\left[\Delta 
N\right]^2\rangle\sim N_s^\beta$ is observed, with $\beta=1$ at low density in 
the homogeneous fluid (ideal gas limit) and $\beta=2$ in the phase separated 
state.
It is known that jammed disordered packings of repulsive particles are 
hyperuniform, i.e., the number fluctuations inside a large window 
grow slower than the volume of the window (in three 
dimensions)~\cite{Donev2005}. This yields a value $\beta<1$ in the frozen state 
obtained at low activity and high density. 
The inset of Figure~\ref{fig:NF} shows the exponent $\beta$ as a 
function of packing fraction at finite activity $\tilde{\vel}=0.1$ (solid line, 
blue online) and its thermal counterpart in the low temperature limit 
$\beta_0=\lim_{\tilde{T}\rightarrow 0}\beta$ with $\tilde{T}=k_BT/(ka^2)$ (dashed line, green online).
The latter clearly displays the hyperuniform nature of number fluctuations,
also apparent in Fig.~\ref{fig:v-MSDexp-NFexp} at low $\tilde{\vel}$.
In fact, the exponent $\beta_0$ characterizing hyperuniformity in the passive limit can also be measured in the active system by taking the limit $\tilde{\vel}\rightarrow0$, yielding similar results but at much greater computational cost.
To account for hyperuniformity, we introduce an effective exponent 
$\beta_{\text{e}}=\beta(\tilde{\vel})-(\beta_0-1)$.
The crossover between a homogeneous fluid and a phase separated state is of course 
smoothed out by finite size effects and we define a phase separated state as one with $\beta_{\text{e}}>\beta_x=1.5$. 
This value is chosen as the one that best matches the results of direct visual observation of the state of the system. 

A surprising finding is that at any fixed $\phi$ above close packing the onset 
of phase separation consistently occurs at a self-propulsion velocity 
$\tilde{\vel}$ well above  the melting from the the frozen state, as defined by the 
long-time behavior of the MSD. This is shown in Fig.~\ref{fig:v-MSDexp-NFexp} 
and supports the existence of a sliver of homogeneous liquid at packing 
fractions above close packing between the frozen and the phase separated 
regions, as shown in Fig.~ \ref{fig:PD_cmap}.

\subsection{Cluster Size Distribution}
\label{csd}

\begin{figure}[h]
\centering
\includegraphics[width=0.9\linewidth]{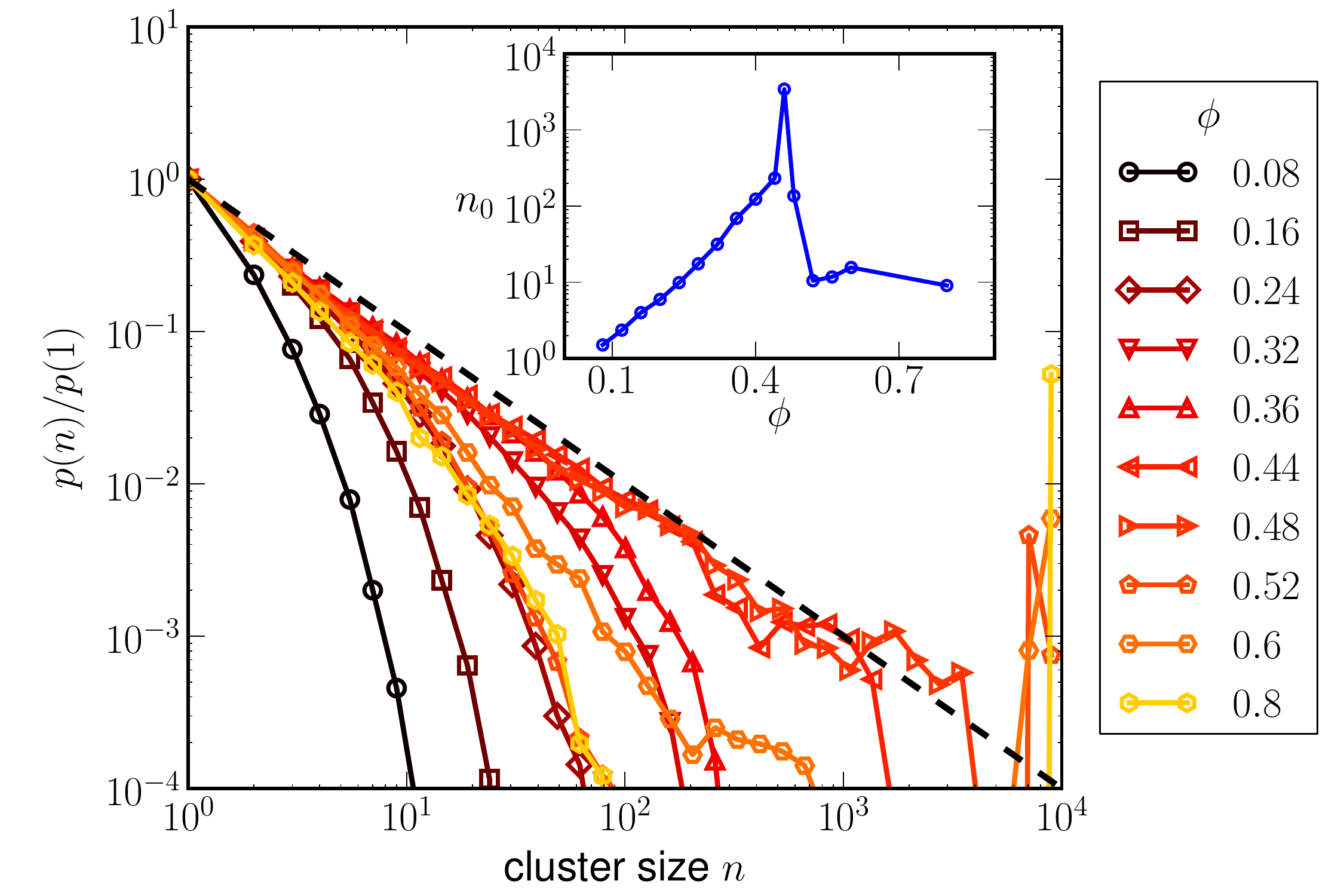}
\caption{
Normalized average fraction $p(n)/p(1)$ of clusters containing $n$ particles for 
a system of $N=10^4$ particles at $\tilde{\vel}=10^{-2}$, 
$\tilde{\nu}_r=5\times10^{-4}$ and various values of $\phi$.
The dashed line is $n^{-1}$.
Inset: cluster size cut-off $n_0$ obtained by fitting $p(n)/p(1)$ to 
$e^{-n/n_0}/n$.
}
\label{fig:CSD}
\end{figure}

On Fig.~\ref{fig:CSD}, we further characterize the system by measuring the cluster size 
distribution (CSD), i.e. the average fraction $p(n)$ of clusters containing $n$ 
particles.
At low density, we find that the CSD is well described by a power law with an 
exponential cut-off $n_0$: $p(n)=p(1)\, e^{-n/n_0}/n$.
The cluster size cut-off $n_0$ increases with density and diverges at the onset 
of phase separation where the CSD reduces to $p(n)\sim n^{-1}$.
In the phase separated region, the CSD exhibits a narrow peak at a value $N_c$ 
equal to the average number of particles in the dense phase while the dilute phase 
contributes the same cut-off power law found in the homogeneous state described 
above.
As density is increased, $N_c$ increases until the dilute phase disappears 
completely ($N_c=N$) and a homogeneous dense phase is obtained, at which point 
$p(n)$ is simply $\delta_{n,N}$ where $\delta$ is the Kronecker symbol.

The presence of the $n^{-1}$ factor in the CSD contrasts with the exponential 
form $p(n)\sim e^{-n/n_0}$ we observe in the thermal case (data not shown)
, and proves that even well below the onset of phase separation, the system cannot be 
described by an effective temperature.
At extreme dilution, when the particles' orientations decorrelate 
much faster than the mean free time ($\zeta \ll 1$, not explored in our simulations), one should recover a thermal-like CSD.

Another striking feature is the extreme slowness of nucleation processes.
Indeed, the onset of phase separation is akin to a first order equilibrium 
transition with distinct coexistence and spinodal curves, and macroscopic 
clusters can form through either nucleation or spinodal decomposition (the 
existence of nucleation-type growth was shown in Ref.~\citeN{Redner2013}, and the 
existence of hysteresis at the transition was confirmed in our simulations in 
the largest systems). Nucleation, however, requires such large seeds and long 
times that one can observe the divergent lengthscale associated with the 
spinodal transition, which would otherwise not be accessible.

Finally, the functional form of the CSD is similar to that obtained in 
Refs.~\citeN{Peruani2006,Peruani2013} by solving mean-field rate equations for 
cluster sizes.
There is however a major difference: 
after the cut-off cluster size diverges, these models exhibit a bimodal CSD but 
no actual phase separation or hysteresis.
Further, this remains true even if the rate equations, originally written for 
aligning particles, are modified to account for the fact that the average speed 
of a cluster decays with its size~\footnote{Assuming particles' orientations are 
random and uncorrelated, the speed of a cluster of size $n$ is $V_n\sim 
n^{-1/2}$, and the relative speed of two clusters of sizes $n_1$, $n_2$ is 
$\Delta V\sim (n_1^{-1}+n_2^{-1})^{1/2}$} (data not shown).

\section{Discussion}

\begin{figure}[h]
\includegraphics[width=0.99\linewidth]{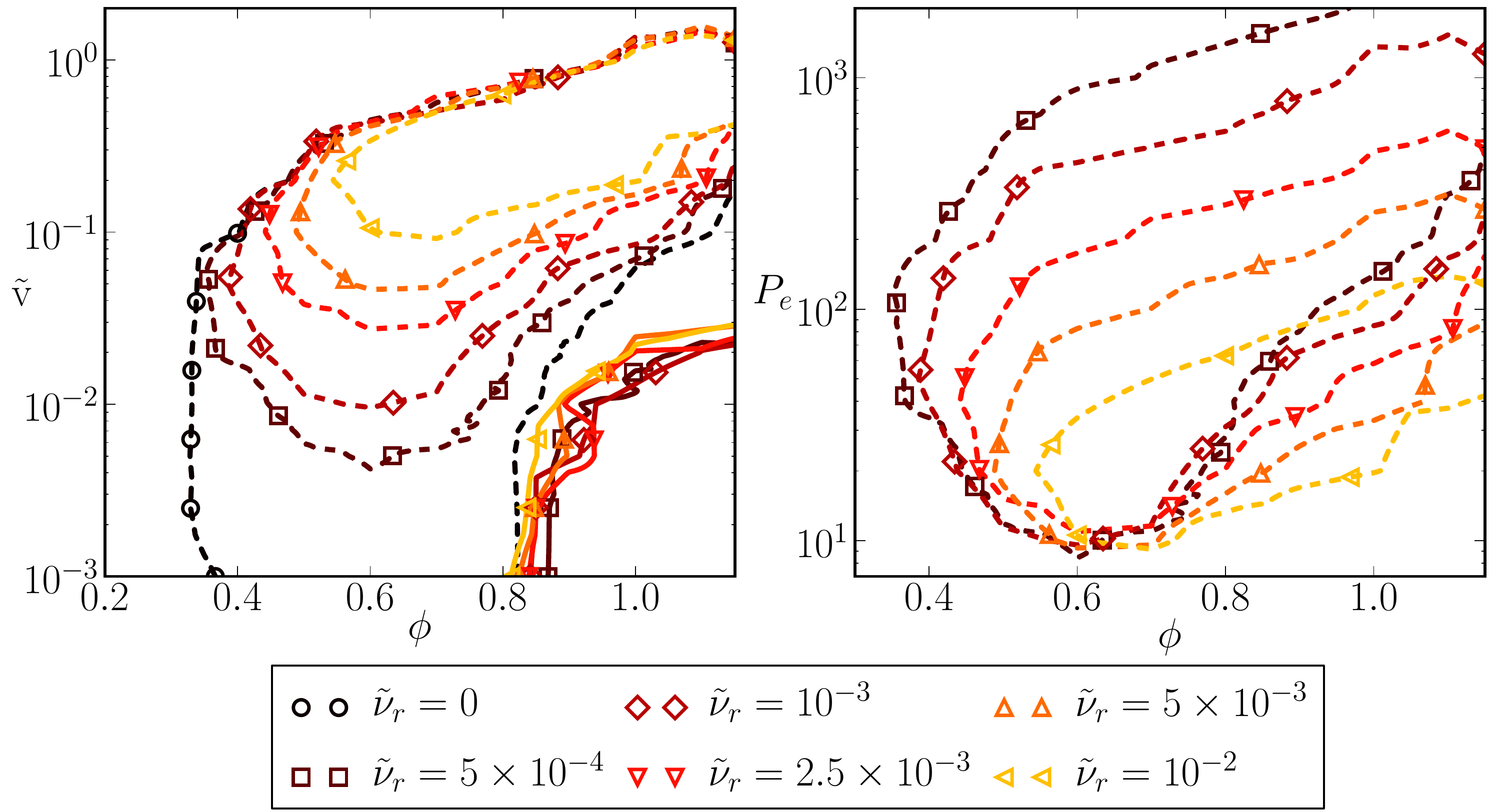}
\caption{
Boundaries of the glassy (solid lines) and phase separated (dashed lines) region 
for various values of the dimensionless angular noise $\tilde{\nu}_r$ in the $\phi-\tilde{\vel}$ plane (left) and in the $\phi-\pe$ plane (right).
The curves marked with squares ($\tilde{\nu}_r=5\times10^{-4}$) are the same as the phase 
boundaries shown on Fig.~\ref{fig:PD_cmap}.
}
\label{fig:PD_outline}
\end{figure}

Several features of the phase diagram deserve comment.
First we show on Fig.~\ref{fig:PD_outline} the outline of the frozen and phase 
separated regions (i.e. the locus of points where $\alpha=\alpha_x$ and 
$\beta_e=\beta_x$, respectively) for various values of the 
dimensionless angular noise $\tilde{\nu}_r$ in the $\phi-\tilde{\vel}$ and $\phi-\pe$ planes.

At high activity ($\tilde{\vel} \geq 1$),  self-propulsion dominates 
over repulsion and our soft disks can pass through each other. The system 
becomes a homogeneous fluid of independent self-propelled particles. 
This crossover clearly depends on the details of the repulsive interaction: it 
is pushed to higher velocities as the particles become stiffer 
\footnote{Stiffness here refers to the force-distance repulsive scaling $F/k\sim r_{ij}^{-x}$ at small $r_{ij}$. Our harmonic potential remains finite, while e.g. $x=12$ for the WCA potential used by Redner et al.~\cite{Redner2013}.}
and should disappear alltogether for infinitely hard particles. 

Conversely, the transition between homogeneous and phase separated fluid at low 
activity has features of a genuine critical point, although more work will be 
needed to establish this.  When plotted in the $(\pe,\phi)$ plane (see right frame of Fig.~\ref{fig:PD_outline}) the phase boundaries seem to collapse on a critical point $\phi^{*}\simeq0.6$ and $\pe^{*}\simeq10$.  
This value is consistent with that reported in Ref.~\citeN{Redner2013} for particles interacting via stiff repulsive interactions
\footnote{
Redner et al. define the P\'{e}clet number as $\pe_R=3\pe$.
That puts our putative critical point at $\pe_R^*\simeq30$, while they observe $\pe_R^*\simeq50$ with WCA interactions and non-zero translational noise\cite{Redner2013}.
}, 
suggesting the onset of phase separation may be 
independent of the form of the repulsive interaction. The existence  of a 
minimal self-propulsion speed $\vel^*$ necessary for phase separation also arises 
from linear stability analysis of the homogeneous 
fluid~\cite{Fily2012,Bialke2013}. 

Another surprising finding is the existence of a minimal packing fraction
below which phase separation never occurs spontaneously, even at infinite P\'eclet number (see $\tilde{\nu}_r=0$ data on the left frame of Fig.~\ref{fig:PD_outline}).
In section ~\ref{mean-field}, we present an estimate for $\vel(\rho)$ that predicts a low density separation boundary
in agreement with this finding. This result is in contrast with the model of Ref.~\citeN{Redner2013}, which 
predicts that the minimal density for cluster stability should decrease as $\pe^{-1}$. 
We stress that both Ref.~\citeN{Redner2013} and Ref.~\citeN{Stenhammar2013}, which do not consider the $\pe$ dependence of $\vel(\rho)$, 
explore a limited range of $\pe\leq 150$, while our results extend to the weakly interacting regime above that number (see Figure~\ref{fig:PD_outline}). Further work is needed to clarify this issue,
and to clearly establish the difference between the binodal envelope mapped numerically by Ref.~\citeN{Redner2013} and 
the spinodal lines predicted here and in Ref.~\citeN{Stenhammar2013}.

Finally, at densities above  random close packing we find a frozen active phase,
consistent with recent results on hard active particles ~\cite{Berthier2013b,Ni2013}.
A surprising observation is the existence of a region of homogeneous liquid 
between the frozen and phase separated regimes,  at packing fractions 
above close packing. 
Both the freezing line and the boundary of the high density separated phase scale linearly with density, with a slope nearly independent of $\tilde{\nu}_r$ in the range of parameters we have explored.

In the following we discuss in more details each of these phenomena and use mean-field theory and scaling arguments to estimate the various boundaries in the phase diagram.

\subsection{Mean-Field Theory and Phase Separation Below Close Packing}
\label{mean-field}

The  dynamics of a homogeneous fluid of self-propelled particles  has been 
described by an effective continuum theory where motility suppression is 
incorporated in a density-dependent propulsion speed 
$\vel(\rho)$~\cite{Tailleur2008,Fily2012,Cates2013,Stenhammar2013}. The mean-field 
model applies when particles experience many collisions before their directed 
motion becomes uncorrelated by rotational noise, i.e., for 
$\zeta>>1$~\cite{Fielding2013}. 
In this regime the large-scale dynamics is described by continuum equations for 
the density $\rho({\bf r},t)$ and the  polarization density field ${\bf p}({\bf 
r},t)$ that describes the local orientation of the particles' axis of 
self-propulsion. Although in the absence of alignment interactions, 
self-propelled disks cannot exhibit a state with orientational order, 
substantial correlations can build up in the local polarization for small values 
of the rotational diffusion rate.  The continuum equations are given 
by~\cite{Fily2012}
\begin{subequations}
\label{hydro}
\begin{gather}
\label{rho}
\partial_t\rho=-\bm\nabla\cdot\left[\vel(\rho){\bf 
p}-D(\rho)\bm\nabla\rho\right]\;,\\
\label{p}
\partial_t{\bf p}=-\nu_r{\bf 
p}-\frac12\bm\nabla\left[\vel(\rho)\rho\right]+K\nabla^2{\bf p}\;,
\end{gather}
\end{subequations}
where for generality we have included a translational diffusion coefficient 
$D(\rho)$ that would arise from thermal noise (not included in our numerics) or 
from interactions.
For times $t\gg\nu_r$, one can neglect the time derivative of the polarization 
in Eq.~\eqref{p} relative to the damping term, solve for ${\bf p}$, and 
eliminate ${\bf p}$ from Eq.~\eqref{rho}. To leading order in the gradients the 
dynamics is then described by a single nonlinear diffusion equation, given by
\begin{align}
\label{diff}
\partial_t\rho=\bm\nabla\cdot\left[{\cal D}(\rho)\bm\nabla\rho\right]\;,
\end{align}
with effective diffusivity 
\begin{align}
\label{Dcal}
{\cal D}(\rho)=D(\rho)+\frac{\vel^2(\rho)}{2\nu_r}\left(1+\frac{d\ln 
\vel}{d\ln\rho}\right)\;.
\end{align}
The linear stability of a homogeneous state of constant density $\rho=\rhobar$ 
can be analyzed by examining the dynamics of density fluctuations, 
$\delta\rho=\rho-\rhobar$, that obey the linear diffusion equation
\begin{align}
\label{diff-lin}
\partial_t\delta\rho={\cal D}(\rhobar)\nabla^2\delta\rho\;.
\end{align}
The vanishing of ${\cal D}(\rhobar)$ signals the instability against the growth 
of density fluctuations. 

To obtain an explicit expression for the instability line, we need an estimate 
for the mean-field self-propulsion speed, $\vel(\rho)$.
At low density, the slowing down of particles is due to the time lost during 
binary collisions. The effective self-propulsion speed can then be written as 
$\vel(\rho)\simeq \vel_0(1-\tau/\tau_c)$ where $\tau_c=1/(2a \vel_0\rho)$ is the 
mean-free time between collisions
and $\tau$ is the average delay caused by each collision.
Considering for simplicity stiff particles (as appropriate for $\tilde{\vel}<<1$), 
we  estimate $\tau$ in two limiting cases.
When particles  maintain their orientation throughout the collision, 
the duration of a collision is simply the time it takes the particles to move 
around each other, or $\tau_1\sim a/\vel_0$.
If, on the other hand, particles reorient during the collision,  the associated 
delay $\tau_2$ is  the reorientation time $\nu_r^{-1}$.
Since the total collision time $\tau$ is controlled by the faster of the two 
mechanisms, we assume additivity of the rates, obtaining 
$\tau^{-1}=\tau_1^{-1}+\tau_2^{-1}$.
The effective velocity is then given by $\vel(\rho)=\vel_0(1-\phi/\phi^*)$ where  
$\phi\simeq \rho/(\pi a^2)$ and $\phi^*(\pe)=2(\phi_1+\phi_2/\pe)$, with 
$\phi_{1,2}$ constants of order $1$ that depend on the detail of the collisions.
Inserting this estimate for $\vel(\rho)$ in the expression for ${\cal D}(\rho)$, and setting 
${\cal D}(\rho)=0$, yields the condition for the linear instability or 
mean-field spinodal line that can be written as
\begin{equation}
\label{eq:spinodal}
{\cal 
D}(\rho) = D(\rho)
+\frac{\vel_0^2}{2\nu_r} \left(1-\phi/\phi^*\right) \left(1-2\phi/\phi^*\right)\;.
\end{equation}
If we neglect $D$, phase separation occurs at 
$\phi_s=\phi^*(\pe)/2=\phi_1+\phi_2/\pe$, which further reduces to the constant 
value $\phi=\phi_1$ at high P\'eclet numbers or in the absence of angular noise. 
In this limit the transition line in the $\phi-\tilde{\vel}$ plane would be 
a vertical line at $\phi=\phi_1$. 
This is indeed what we observe when $\tilde{\nu}_r=0$ (see left frame of Fig.~\ref{fig:PD_outline}).
Above $\tilde{\vel}\approx0.1$ particle overlap 
becomes, however, significant  and suppresses phase separation. A fit to the 
simple estimate $\phi_s=\phi^*(\pe)$ is shown in Fig.~\ref{fig:PD_cmap} as a 
dotted line. We note that 
that close to the putative critical point, when the $\pe^{-1}$ term in $\phi^*$ 
becomes dominant, our expression for the spinodal line reduces (up to numerical 
factors) to that obtained in Ref.~\citeN{Redner2013} for the coexistence line via 
an analysis of the stability of a gas-cluster interface.
This is not surprising as one expects the spinodal and the binodal lines to 
merge at the critical point.

A finite value of $D$ yields a lower bound to the value of $\pe$ required for 
the onset of phase separation, as discussed earlier~\cite{Fily2012,Bialke2013}. 
This can be seen by neglecting the density dependence of $D(\rho)$.
Eq.~\eqref{eq:spinodal} is then easily solved for the spinodal boundary, with 
the result  
\begin{equation}
\label{eq:phis}
\phi_{\text{s}}=\frac{\phi^*(\pe)}{4}\left(3+\sqrt{1-\left(\frac{\pe_m}{\pe}
\right)^2}\right)\;,
\end{equation}
where $\pe_m=4\sqrt{D/(a^2\nu_r)}$ represents the  a minimum value of $\pe$ required for phase separation.

For finite orientational persistence time $\nu_r^{-1}$, we need to retain the 
dynamics of polarization. The linear stability of the homogeneous state can then 
be examined by linearizing Eqs.~\eqref{hydro} around $\rho=\rhobar$ and ${\bf 
p}=0$. Working in Fourier space, we let $\left\{\delta\rho({\bf r},t),{\bf 
p}({\bf r},t)\right\}=\sum_{\bf q}\left\{\delta\rho_{\bf q}(t),{\bf p}_{\bf 
q}(t)\right\}e^{i{\bf q}\cdot{\bf r}}$. The time evolution of the Fourier 
amplitudes is governed by the coupled equations
\begin{align}
& \left(\partial_t+Dq^2\right)\delta\rho_{\bf q}+i\vel{\bf q}\cdot{\bf p}_{\bf 
q}=0\;,\\
& \left(\partial_t+\nu_r+Kq^2\right){\bf p}_{\bf q}+\frac12(\vel+\rhobar \vel')i{\bf 
q}\delta\rho_{\bf q}=0\;.
\label{eq:linear}
\end{align}
where $\vel'=d\vel/d\rho$ and all quantities are evaluated at the homogeneous density 
$\rhobar$. The decay/growth of fluctuations is governed by the eigenvalues of 
Eqs.~\eqref{eq:linear}. The stability is controlled by the eigenvalue $s_+(q)$, 
with dispersion relation
\begin{align}
\label{eq:sp}
s_+(q) = & -\frac12\left[\nu_r+(K+D)q^2\right]\notag\\
       & +\frac12\sqrt{\left[\nu_r+(K-D)q^2\right]^2-2q^2\vel(\vel+\rhobar \vel')}\;.
\end{align}
\begin{figure}[h]
 \centering
 \includegraphics[width=0.8\linewidth]{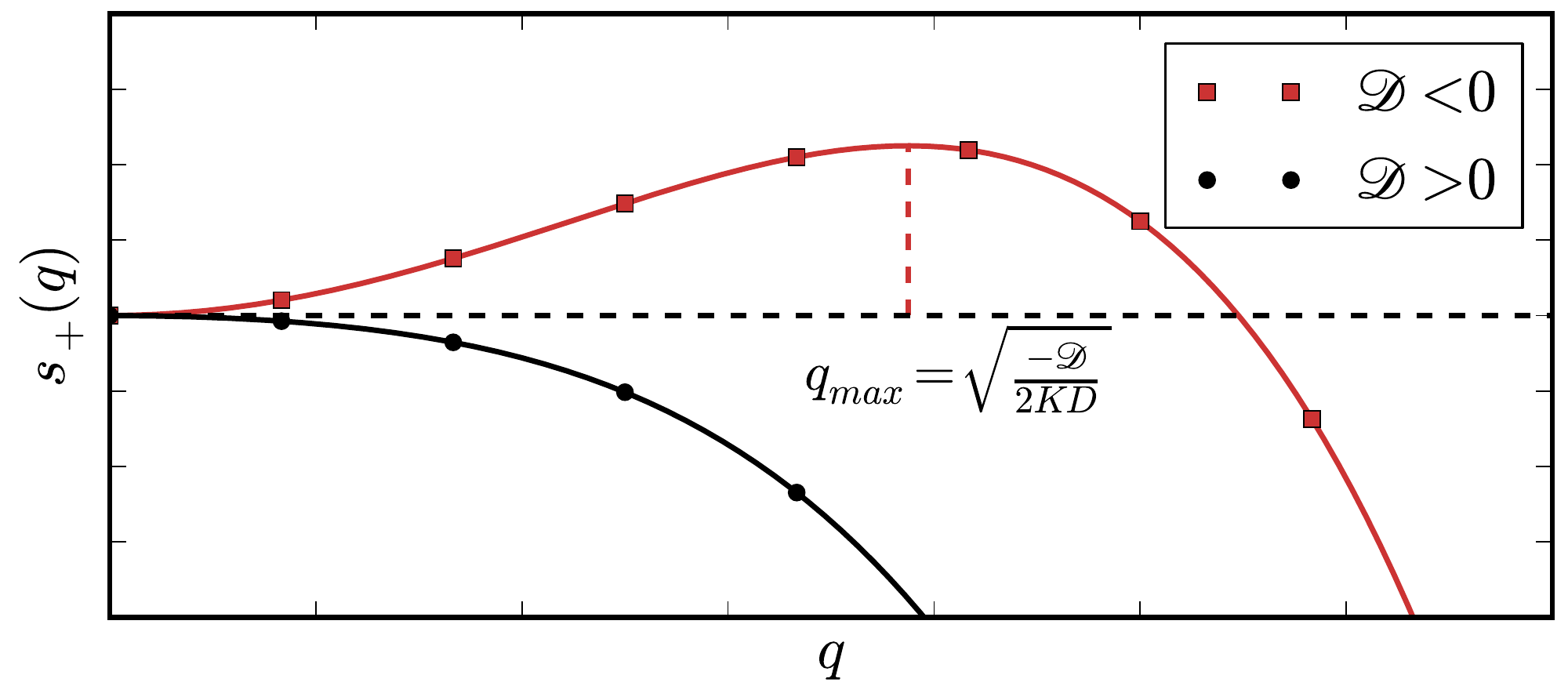}
 \caption{Schematic dispersion relation $s_{+}(q)$ for ${\cal D}>0$ and ${\cal D}<0$.}
 \label{fig:dispersion}
\end{figure}

For small wave vectors, the dispersion relation can be approximated as 
$s_+(q)\simeq -q^2{\cal D}(\rhobar)+q^4B$, with $B={4\nu_r^2}\vel(\vel+\rhobar 
\vel')\left[2(K-D)-\vel(\vel+\rhobar \vel')/\nu_r\right]$.  The rate $s_+$ becomes positive 
for ${\cal D}<0$, indicating the growth of a sinusoidal modulation of the 
density. The growth rate as a function of the wavevector $q$  is shown in Fig.~\ref{fig:dispersion} and has 
the typical behavior expected in a spinodal regime. All modulations with 
$q<q_0=\sqrt{-{\cal D}/B}$ are unstable. The maximum growth rate is obtained at 
$q_{max}=q_0/\sqrt{2}$. At the onset of the instability modes of wavevector 
$q_{max}\sim \sqrt{-{\cal D}/(2KD)}$  will dominate experimental observations.

\subsection{Freezing and Phase Separation Above Close Packing}
\label{above}
We observe an arrested phase at high density and low self propulsion speed. 
While the existence of dynamically arrested states of active systems has been 
reported before 
~\cite{Henkes2011,bialke_crystallization_2012,Redner2013,
reichhardt_dynamical_2011}, the dynamics of the freezing transition has only started to attract interest 
recently~\cite{Berthier2013a,Berthier2013b,Ni2013}.

Berthier~\cite{Berthier2013b} and Ni~\cite{Ni2013} have reported  a shift of the 
glass transition of active particles to packing fractions higher than those 
known for thermal systems, in both two and three dimensions. The shift grows 
with a parameter controlling the persistence time of the single-particle 
dynamics, essentially our $\nu_r^{-1}$. Both simulations consider polydisperse 
\emph{hard} particles, which differ from our soft-disk system in that they cannot be compressed beyond the random close packing  
(RCP)  fraction, $\phi_{RCP} \approx 0.842$ in two dimensions.
On the other hand, below RCP and in the limit $k\rightarrow\infty$,
the interparticle overlap for soft disks vanishes and hard and soft particles should exhibit similar behavior. More precisely, for $\phi<\phi_{RCP}$ passive soft disks will behave like hard ones provided the interaction time scale $(\mu k)^{-1}$ is short compared to the persistence time $\nu_r^{-1}$, 
i.e.,  $\tilde{\nu}_r=\nu_r/(\mu k)\ll1$. We then expect active soft disks to behave like hard ones provided $\tilde{\nu}_r\ll1$ and $\tilde{\vel}=\vel_0/(a\mu k)\ll1$. 
From  Fig.~\ref{fig:PD_cmap}, we extract the onset of a glassy state in the limit $\tilde{\vel}\rightarrow 0$ at $\phi_G(\tilde{\vel}\rightarrow 0)=0.86 \pm 0.02 $ for $\tilde{\nu}_r=5\times 10^{-4}$, 
above the thermal $\phi_{G}\approx 0.8$ and consistent with the value close to $\phi_{RCP}$ reported by Berthier~\cite{Berthier2013b}. It is known that athermal homogeneously sheared systems jam at $\phi_{RCP}>\phi_G$ in the limit of vanishing strain rate~\cite{Olsson2007,Ikeda2012}. 
One can then speculate that activity, in spite of injecting energy locally rather than globally,  be equivalent to a homogeneous shear in the limit $\tilde{\nu}_r\rightarrow 0$.
Finally, we stress  that a shift of the kinetic arrest to values of packing fractions above the glass transition was also reported for self-propelled disks with aligning interactions~\cite{Henkes2011}. 

At finite $\vel_0$,  the glass transition line shifts to higher packing fractions (Fig.~\ref{fig:PD_cmap}). 
In addition, the phase separated state still exists at high $\vel_0$ and packing fraction above $\phi_{RCP}$, but is separated from the glassy phase by an intermediate region of homogeneous liquid. 
This gap between glass and phase separated state for $\phi>\phi_{RCP}$ is  perhaps the most surprising feature of the phase diagrams in Figs.~\ref{fig:PD_cmap} and~\ref{fig:PD_outline}. 
Inspection of the dynamics in this region clearly shows no significant clustering and a substantial amount of local rearrangements, indicative of a fluid phase (see supplementary movie 1(d)).
The gap is also apparent from  Fig.~\ref{fig:v-MSDexp-NFexp} showing the behavior of the exponents  $\alpha$ (MSD) and $\beta$ (number fluctuations) along vertical (i.e. constant $\phi$) slices of the phase diagram of 
Fig.~\ref{fig:PD_cmap}, above close packing. 
Both the glass transition line and the phase separation at high density can be understood as arising from a balance between the passive pressure of a densely packed system and the active forces due to self propulsion. 
In the absence of activity, the pressure of passive soft disks repelling via one-sided harmonic springs of stiffness $k$ scales linearly  with the particle overlap $\delta$, 
which is positive at packing fractions above RCP,  i.e., $p=k\delta$. 
Even though the packings are disordered and rearrangements are non-affine, the athermal equation 
of state in the solid phase remains linear with $p=p_0(\phi-\phi_{\text{RCP}})$ and $p_0=0.34k$~\cite{O'Hern2003}.
In the active system, a pressure $p_a\approx \vel_0/2 a \mu$ arises from self-propelled particles pushing against their neighbors. 
Balancing these two pressures yields a critical velocity in dimensionless units 
$\tilde{\vel}^*=\vel^*/(a\mu k) \sim u (\phi - \phi_{RCP})$, with $u$ a dimensionless parameter that should be compared to the full pressure balance at $u^{*}=2p_0/k=0.68$.
The glass transition line can be fitted by this expression with $u=0.07$, 
suggestive of a simple melting criterion where active particles squeeze through gaps between their neighbors at $\sim10\%$ of the passive interparticle pressure. 
We stress that in the range of $\tilde{\nu}_r$ we explored ($\tilde{\nu}_r=0$ to $10^{-2}$), the glass transition  line is indeed
nearly independent of $\tilde{\nu}_r$ and depends only on $\phi$. 

In contrast, the boundary between the homogeneous liquid and the phase separated state
at packing fractions above $\phi_{RCP}$ depends on both $\phi$ and $\tilde{\nu}_r$. 
When $\tilde{\nu}_r\rightarrow 0$, this boundary line and the glass transition line converge 
at $\phi_{RCP}$ when $\tilde{\vel}\rightarrow 0$ (see Fig.~\ref{fig:PD_outline}).
The boundary line may again be interpreted as a pressure balance, this time at the interface that arises in the phase separated state between the dense liquid and the gas.
First, we note that the surface of the dense phase consists of a highly polarized layer of particles pointing inwards. Indeed, particles pointing away from the dense phase immediately leave it%
\footnote{This is also the basis for the interface stability argument in Ref.~\citeN{Redner2013}.}.
The active pressure exerted by this layer on the dense phase is $p_a\sim \gamma \vel_0/(2a\mu)$
where $\vel_0/\mu$ is the magnitude of the active force exerted by a single particle and $2a$ is the average distance between particles at the interface. 
The dimensionless factor $\gamma$, of order $1$, captures information about the distribution of orientations in the interfacial layer and its thickness 
($\gamma=1$ for a single layer of particles all oriented normal to the surface).
The passive pressure in the dense phase, on the other hand, is at least $p=p_0(\phi-\phi_{RCP})$ with $\phi$ the average packing fraction in the system.
Balancing the two yields again a critical velocity $\tilde{\vel}^* = u (\phi - \phi_{RCP})$, where $u$ should now be compared to $u^{*}/\gamma$. 
It fits the phase separation boundary at $\tilde{\nu}_r=0$ for $u=0.7$, indicating that $\gamma \approx 1$.
While melting requires overcoming only about $10\%$ of the passive pressure, phase separation occurs when activity is large enough to fully balance the passive pressure due to elastic forces. 
For $\tilde{\nu}_r>0$, the gap for the onset of phase separation can be accounted for by using $\tilde{\vel}^*=u(\phi-\phi_{RCP})+w(\tilde{\nu}_r)$. 
The fit gives $u=0.7$, independent of $\tilde{\nu}_r$ and $w\sim\tilde{\nu}_r^{1/2}$ at small $\tilde{\nu}_r$.
More work is needed to understand whether the existence of a gap between melting and phase separation is generic for soft interparticle potentials.

\section{Summary}
In this article, we present a numerical characterization of the phase diagram of self-propelled particles with only repulsive interactions and no alignment
that shows  the existence of three distinct phases in the system: a homogeneous liquid, a phase-separated liquid, and a glassy state. We show evidence that  the phase separated region is bounded at low P\'eclet number by a critical point (Fig.~\ref{fig:PD_outline}), including scaling arguments for the spinodal line at both low and high packing fraction, consistent with the idea of a first order phase transition. 
Analysis of the CSD shows that the typical cluster size diverges as the spinodal line is approached, while non-thermal effects persist well below the transition.
We also show the existence of a frozen phase at high packing fraction
and low active self-propulsion and find a novel high density active
liquid phase in the gap between the frozen and phase separated states. 
More work is required to provide a detailed characterization of the critical point at low $\pe$,
in particular to differentiate spinodal and binodal lines taking nucleation and finite size effects into account.
The existence of an active glass or jamming transition different from both its thermal and sheared counterparts
is an intriguing result that deserves further exploration.

\begin{acknowledgments}
MCM acknowledges support of the the National Science Foundation through awards DMR-1004789 and DGE-1068780. The computations were carried out on the Syracuse University HTC Campus Grid.
\end{acknowledgments}

\bibliography{biblio}

\begin{thebibliography}{10}%
\makeatletter
\providecommand \@ifxundefined [1]{%
 \ifx #1\undefined \expandafter \@firstoftwo
 \else \expandafter \@secondoftwo
\fi
}%
\providecommand \@ifnum [1]{%
 \ifnum #1\expandafter \@firstoftwo
 \else \expandafter \@secondoftwo
\fi
}%
\providecommand \enquote [1]{``#1''}%
\providecommand \bibnamefont  [1]{#1}%
\providecommand \bibfnamefont [1]{#1}%
\providecommand \citenamefont [1]{#1}%
\providecommand\href[0]{\@sanitize\@href}%
\providecommand\@href[1]{\endgroup\@@startlink{#1}\endgroup\@@href}%
\providecommand\@@href[1]{#1\@@endlink}%
\providecommand \@sanitize [0]{\begingroup\catcode`\&12\catcode`\#12\relax}%
\@ifxundefined \pdfoutput {\@firstoftwo}{%
 \@ifnum{\z@=\pdfoutput}{\@firstoftwo}{\@secondoftwo}%
}{%
 \providecommand\@@startlink[1]{\leavevmode\special{html:<a href="#1">}}%
 \providecommand\@@endlink[0]{\special{html:</a>}}%
}{%
 \providecommand\@@startlink[1]{%
  \leavevmode
  \pdfstartlink
   attr{/Border[0 0 1 ]/H/I/C[0 1 1]}%
   user{/Subtype/Link/A<</Type/Action/S/URI/URI(#1)>>}%
  \relax
 }%
 \providecommand\@@endlink[0]{\pdfendlink}%
}%
\providecommand \url  [0]{\begingroup\@sanitize \@url }%
\providecommand \@url [1]{\endgroup\@href {#1}{\urlprefix}}%
\providecommand \urlprefix [0]{URL }%
\providecommand \Eprint[0]{\href }%
\@ifxundefined \urlstyle {%
  \providecommand \doi [1]{doi:\discretionary{}{}{}#1}%
}{%
  \providecommand \doi [0]{doi:\discretionary{}{}{}\begingroup
  \urlstyle{rm}\Url }%
}%
\providecommand \doibase [0]{http://dx.doi.org/}%
\providecommand \Doi[1]{\href{\doibase#1}}%
\providecommand \bibAnnote [3]{%
  \BibitemShut{#1}%
  \begin{quotation}\noindent
    \textsc{Key:}\ #2\\\textsc{Annotation:}\ #3%
  \end{quotation}%
}%
\providecommand \bibAnnoteFile [2]{%
  \IfFileExists{#2}{\bibAnnote {#1} {#2} {\input{#2}}}{}%
}%
\providecommand \typeout [0]{\immediate \write \m@ne }%
\providecommand \selectlanguage [0]{\@gobble}%
\providecommand \bibinfo [0]{\@secondoftwo}%
\providecommand \bibfield [0]{\@secondoftwo}%
\providecommand \translation [1]{[#1]}%
\providecommand \BibitemOpen[0]{}%
\providecommand \bibitemStop [0]{}%
\providecommand \bibitemNoStop [0]{.\EOS\space}%
\providecommand \EOS [0]{\spacefactor3000\relax}%
\providecommand \BibitemShut [1]{\csname bibitem#1\endcsname}%
\bibitem{Marchetti2013}%
  \BibitemOpen
  \bibfield{author}{%
  \bibinfo {author} {\bibfnamefont{M.~C.}\ \bibnamefont{Marchetti}}, \bibinfo
  {author} {\bibfnamefont{J.~F.}\ \bibnamefont{Joanny}}, \bibinfo {author}
  {\bibfnamefont{S.~R.}\ \bibnamefont{Ramaswamy}}, \bibinfo {author}
  {\bibfnamefont{T.~B.}\ \bibnamefont{Liverpool}}, \bibinfo {author}
  {\bibfnamefont{J.}~\bibnamefont{Prost}}, \bibinfo {author}
  {\bibfnamefont{M.}~\bibnamefont{Rao}},\ and\ \bibinfo {author}
  {\bibfnamefont{R.~A.}\ \bibnamefont{Simha}},\ }%
  \bibfield{journal}{%
  \bibinfo {journal} {Rev. Mod. Phys.}\ }%
  \textbf{\bibinfo {volume} {85}},\ \bibinfo {pages} {1143} (\bibinfo {year}
  {2013}),\ \url{http://arxiv.org/abs/1207.2929}%
  \bibAnnoteFile{NoStop}{Marchetti2013}%
\bibitem{Schaller2011}%
  \BibitemOpen
  \bibfield{author}{%
  \bibinfo {author} {\bibfnamefont{V.}~\bibnamefont{Schaller}}, \bibinfo
  {author} {\bibfnamefont{C.~A.}\ \bibnamefont{Weber}}, \bibinfo {author}
  {\bibfnamefont{B.}~\bibnamefont{Hammerich}}, \bibinfo {author}
  {\bibfnamefont{E.}~\bibnamefont{Frey}},\ and\ \bibinfo {author}
  {\bibfnamefont{A.~R.}\ \bibnamefont{Bausch}},\ }%
  \bibfield{journal}{%
  \Doi{10.1073/pnas.1107540108}{\bibinfo {journal} {Proceedings of the National
  Academy of Sciences}}\ }%
  \textbf{\bibinfo {volume} {108}},\ \bibinfo {pages} {19183 } (\bibinfo
  {month} {Nov.}\ \bibinfo {year} {2011}),\
  \url{http://www.pnas.org/content/108/48/19183.abstract}%
  \bibAnnoteFile{NoStop}{Schaller2011}%
\bibitem{peruani2012}%
  \BibitemOpen
  \bibfield{author}{%
  \bibinfo {author} {\bibfnamefont{F.}~\bibnamefont{Peruani}}, \bibinfo
  {author} {\bibfnamefont{J.}~\bibnamefont{Starru{\ss}}}, \bibinfo {author}
  {\bibfnamefont{V.}~\bibnamefont{Jakovljevic}}, \bibinfo {author}
  {\bibfnamefont{L.}~\bibnamefont{S{\o}gaard-Andersen}}, \bibinfo {author}
  {\bibfnamefont{A.}~\bibnamefont{Deutsch}},\ and\ \bibinfo {author}
  {\bibfnamefont{M.}~\bibnamefont{B\"ar}},\ }%
  \bibfield{journal}{%
  \Doi{10.1103/PhysRevLett.108.098102}{\bibinfo {journal} {Phys. Rev. Lett.}}\
  }%
  \textbf{\bibinfo {volume} {108}},\ \bibinfo {pages} {098102} (\bibinfo
  {month} {Feb.}\ \bibinfo {year} {2012})%
  \bibAnnoteFile{NoStop}{peruani2012}%
\bibitem{Poujade2007}%
  \BibitemOpen
  \bibfield{author}{%
  \bibinfo {author} {\bibfnamefont{M.}~\bibnamefont{Poujade}}, \bibinfo
  {author} {\bibfnamefont{E.}~\bibnamefont{Grasland-Mongrain}}, \bibinfo
  {author} {\bibfnamefont{A.}~\bibnamefont{Hertzog}}, \bibinfo {author}
  {\bibfnamefont{J.}~\bibnamefont{Jouanneau}}, \bibinfo {author}
  {\bibfnamefont{P.}~\bibnamefont{Chavrier}}, \bibinfo {author}
  {\bibfnamefont{B.}~\bibnamefont{Ladoux}}, \bibinfo {author}
  {\bibfnamefont{A.}~\bibnamefont{Buguin}},\ and\ \bibinfo {author}
  {\bibfnamefont{P.}~\bibnamefont{Silberzan}},\ }%
  \bibfield{journal}{%
  \bibinfo {journal} {PNAS}\ }%
  \textbf{\bibinfo {volume} {104}},\ \bibinfo {pages} {15988} (\bibinfo {year}
  {2007})%
  \bibAnnoteFile{NoStop}{Poujade2007}%
\bibitem{Trepat2009}%
  \BibitemOpen
  \bibfield{author}{%
  \bibinfo {author} {\bibfnamefont{X.}~\bibnamefont{Trepat}}, \bibinfo {author}
  {\bibfnamefont{M.~R.}\ \bibnamefont{Wasserman}}, \bibinfo {author}
  {\bibfnamefont{T.~E.}\ \bibnamefont{Angelini}}, \bibinfo {author}
  {\bibfnamefont{E.}~\bibnamefont{Millet}}, \bibinfo {author}
  {\bibfnamefont{D.~A.}\ \bibnamefont{Weitz}}, \bibinfo {author}
  {\bibfnamefont{J.~P.}\ \bibnamefont{Butler}},\ and\ \bibinfo {author}
  {\bibfnamefont{J.~J.}\ \bibnamefont{Fredberg}},\ }%
  \bibfield{journal}{%
  \bibinfo {journal} {Nat. Phys.}\ }%
  \textbf{\bibinfo {volume} {5}},\ \bibinfo {pages} {426} (\bibinfo {year}
  {2009})%
  \bibAnnoteFile{NoStop}{Trepat2009}%
\bibitem{Ballerini2008}%
  \BibitemOpen
  \bibfield{author}{%
  \bibinfo {author} {\bibfnamefont{M.}~\bibnamefont{Ballerini}}, \bibinfo
  {author} {\bibfnamefont{N.}~\bibnamefont{Cabibbo}}, \bibinfo {author}
  {\bibfnamefont{R.}~\bibnamefont{Candelier}}, \bibinfo {author}
  {\bibfnamefont{A.}~\bibnamefont{Cavagna}}, \bibinfo {author}
  {\bibfnamefont{E.}~\bibnamefont{Cisbani}}, \bibinfo {author}
  {\bibfnamefont{I.}~\bibnamefont{Giardina}}, \bibinfo {author}
  {\bibfnamefont{V.}~\bibnamefont{Lecomte}}, \bibinfo {author}
  {\bibfnamefont{A.}~\bibnamefont{Orlandi}}, \bibinfo {author}
  {\bibfnamefont{G.}~\bibnamefont{Parisi}}, \bibinfo {author}
  {\bibfnamefont{A.}~\bibnamefont{Procaccini}}, \bibinfo {author}
  {\bibfnamefont{M.}~\bibnamefont{Viale}},\ and\ \bibinfo {author}
  {\bibfnamefont{V.}~\bibnamefont{Zdravkovic}},\ }%
  \bibfield{journal}{%
  \bibinfo {journal} {Proceedings of the National Academy of Sciences}\ }%
  \textbf{\bibinfo {volume} {105}},\ \bibinfo {pages} {1232} (\bibinfo {month}
  {Jan.}\ \bibinfo {year} {2008}),\ ISSN \bibinfo {issn} {0027-8424,
  1091-6490}%
  \bibAnnoteFile{NoStop}{Ballerini2008}%
\bibitem{Palacci2010}%
  \BibitemOpen
  \bibfield{author}{%
  \bibinfo {author} {\bibfnamefont{J.}~\bibnamefont{Palacci}}, \bibinfo
  {author} {\bibfnamefont{B.}~\bibnamefont{Ab\'{e}cassis}}, \bibinfo {author}
  {\bibfnamefont{C.}~\bibnamefont{Cottin-Bizonne}}, \bibinfo {author}
  {\bibfnamefont{C.}~\bibnamefont{Ybert}},\ and\ \bibinfo {author}
  {\bibfnamefont{L.}~\bibnamefont{Bocquet}},\ }%
  \bibfield{journal}{%
  \Doi{10.1103/PhysRevLett.104.138302}{\bibinfo {journal} {Physical Review
  Letters}}\ }%
  \textbf{\bibinfo {volume} {104}},\ \bibinfo {pages} {138302} (\bibinfo
  {month} {Apr.}\ \bibinfo {year} {2010}),\
  \url{http://link.aps.org/doi/10.1103/PhysRevLett.104.138302}%
  \bibAnnoteFile{NoStop}{Palacci2010}%
\bibitem{Theurkauff2012}%
  \BibitemOpen
  \bibfield{author}{%
  \bibinfo {author} {\bibfnamefont{I.}~\bibnamefont{Theurkauff}}, \bibinfo
  {author} {\bibfnamefont{C.}~\bibnamefont{Cottin-Bizonne}}, \bibinfo {author}
  {\bibfnamefont{J.}~\bibnamefont{Palacci}}, \bibinfo {author}
  {\bibfnamefont{C.}~\bibnamefont{Ybert}},\ and\ \bibinfo {author}
  {\bibfnamefont{L.}~\bibnamefont{Bocquet}},\ }%
  \bibfield{journal}{%
  \Doi{10.1103/PhysRevLett.108.268303}{\bibinfo {journal} {Physical Review
  Letters}}\ }%
  \textbf{\bibinfo {volume} {108}},\ \bibinfo {pages} {268303} (\bibinfo
  {month} {Jun.}\ \bibinfo {year} {2012}),\
  \url{http://link.aps.org/doi/10.1103/PhysRevLett.108.268303}%
  \bibAnnoteFile{NoStop}{Theurkauff2012}%
\bibitem{Thutupalli2011}%
  \BibitemOpen
  \bibfield{author}{%
  \bibinfo {author} {\bibfnamefont{S.}~\bibnamefont{Thutupalli}}, \bibinfo
  {author} {\bibfnamefont{R.}~\bibnamefont{Seemann}},\ and\ \bibinfo {author}
  {\bibfnamefont{S.}~\bibnamefont{Herminghaus}},\ }%
  \bibfield{journal}{%
  \Doi{10.1088/1367-2630/13/7/073021}{\bibinfo {journal} {New Journal of
  Physics}}\ }%
  \textbf{\bibinfo {volume} {13}},\ \bibinfo {pages} {073021} (\bibinfo {month}
  {Jul.}\ \bibinfo {year} {2011}),\ ISSN \bibinfo {issn} {1367-2630},\
  \url{http://iopscience.iop.org/1367-2630/13/7/073021}%
  \bibAnnoteFile{NoStop}{Thutupalli2011}%
\bibitem{Sanchez2012}%
  \BibitemOpen
  \bibfield{author}{%
  \bibinfo {author} {\bibfnamefont{T.}~\bibnamefont{Sanchez}}, \bibinfo
  {author} {\bibfnamefont{D.~T.~N.}\ \bibnamefont{Chen}}, \bibinfo {author}
  {\bibfnamefont{S.~J.}\ \bibnamefont{DeCamp}}, \bibinfo {author}
  {\bibfnamefont{M.}~\bibnamefont{Heymann}},\ and\ \bibinfo {author}
  {\bibfnamefont{Z.}~\bibnamefont{Dogic}},\ }%
  \bibfield{journal}{%
  \bibinfo {journal} {Nature}\ }%
  \textbf{\bibinfo {volume} {491}},\ \bibinfo {pages} {431} (\bibinfo {month}
  {Nov.}\ \bibinfo {year} {2012}),\ ISSN \bibinfo {issn} {0028-0836},\
  \url{http://dx.doi.org/10.1038/nature11591}%
  \bibAnnoteFile{NoStop}{Sanchez2012}%
\bibitem{Deseigne2010}%
  \BibitemOpen
  \bibfield{author}{%
  \bibinfo {author} {\bibfnamefont{J.}~\bibnamefont{Deseigne}}, \bibinfo
  {author} {\bibfnamefont{O.}~\bibnamefont{Dauchot}},\ and\ \bibinfo {author}
  {\bibfnamefont{H.}~\bibnamefont{Chat\'e}},\ }%
  \bibfield{journal}{%
  \Doi{10.1103/PhysRevLett.105.098001}{\bibinfo {journal} {Phys. Rev. Lett.}}\
  }%
  \textbf{\bibinfo {volume} {105}},\ \bibinfo {pages} {098001} (\bibinfo
  {month} {Aug.}\ \bibinfo {year} {2010})%
  \bibAnnoteFile{NoStop}{Deseigne2010}%
\bibitem{Fily2012}%
  \BibitemOpen
  \bibfield{author}{%
  \bibinfo {author} {\bibfnamefont{Y.}~\bibnamefont{Fily}}\ and\ \bibinfo
  {author} {\bibfnamefont{M.~C.}\ \bibnamefont{Marchetti}},\ }%
  \bibfield{journal}{%
  \Doi{10.1103/PhysRevLett.108.235702}{\bibinfo {journal} {Physical Review
  Letters}}\ }%
  \textbf{\bibinfo {volume} {108}},\ \bibinfo {pages} {235702} (\bibinfo
  {month} {Jun.}\ \bibinfo {year} {2012}),\
  \url{http://link.aps.org/doi/10.1103/PhysRevLett.108.235702}%
  \bibAnnoteFile{NoStop}{Fily2012}%
\bibitem{Redner2013}%
  \BibitemOpen
  \bibfield{author}{%
  \bibinfo {author} {\bibfnamefont{G.~S.}\ \bibnamefont{Redner}}, \bibinfo
  {author} {\bibfnamefont{M.~F.}\ \bibnamefont{Hagan}},\ and\ \bibinfo {author}
  {\bibfnamefont{A.}~\bibnamefont{Baskaran}},\ }%
  \bibfield{journal}{%
  \bibinfo {journal} {Phys. Rev. Lett.}\ }%
  \textbf{\bibinfo {volume} {110}},\ \bibinfo {pages} {055701} (\bibinfo
  {month} {Jan.}\ \bibinfo {year} {2013}),\
  \url{http://link.aps.org/doi/10.1103/PhysRevLett.110.055701}%
  \bibAnnoteFile{NoStop}{Redner2013}%
\bibitem{Cates2012}%
  \BibitemOpen
  \bibfield{author}{%
  \bibinfo {author} {\bibfnamefont{M.~E.}\ \bibnamefont{Cates}},\ }%
  \bibfield{journal}{%
  \Doi{10.1088/0034-4885/75/4/042601}{\bibinfo {journal} {Reports on Progress
  in Physics}}\ }%
  \textbf{\bibinfo {volume} {75}},\ \bibinfo {pages} {042601} (\bibinfo {year}
  {2012}),\ ISSN \bibinfo {issn} {0034-4885}%
  \bibAnnoteFile{NoStop}{Cates2012}%
\bibitem{Cates2013}%
  \BibitemOpen
  \bibfield{author}{%
  \bibinfo {author} {\bibfnamefont{M.~E.}\ \bibnamefont{Cates}}\ and\ \bibinfo
  {author} {\bibfnamefont{J.}~\bibnamefont{Tailleur}},\ }%
  \bibfield{journal}{%
  \bibinfo {journal} {EPL (Europhysics Letters)}\ }%
  \textbf{\bibinfo {volume} {101}},\ \bibinfo {pages} {20010} (\bibinfo {year}
  {2013}),\ ISSN \bibinfo {issn} {0295-5075}%
  \bibAnnoteFile{NoStop}{Cates2013}%
\bibitem{Stenhammar2013}%
  \BibitemOpen
  \bibfield{author}{%
  \bibinfo {author} {\bibfnamefont{J.}~\bibnamefont{Stenhammar}}, \bibinfo
  {author} {\bibfnamefont{A.}~\bibnamefont{Tiribocchi}}, \bibinfo {author}
  {\bibfnamefont{R.~J.}\ \bibnamefont{Allen}}, \bibinfo {author}
  {\bibfnamefont{D.}~\bibnamefont{Marenduzzo}},\ and\ \bibinfo {author}
  {\bibfnamefont{M.~E.}\ \bibnamefont{Cates}},\ }%
  \enquote{\bibinfo {title} {A continuum theory of phase separation kinetics
  for active brownian particles},}\  (\bibinfo {year} {2013}),\
  \url{http://arxiv.org/abs/1307.4373}%
  \bibAnnoteFile{NoStop}{Stenhammar2013}%
\bibitem{bialke_crystallization_2012}%
  \BibitemOpen
  \bibfield{author}{%
  \bibinfo {author} {\bibfnamefont{J.}~\bibnamefont{Bialk\'e}}, \bibinfo
  {author} {\bibfnamefont{T.}~\bibnamefont{Speck}},\ and\ \bibinfo {author}
  {\bibfnamefont{H.}~\bibnamefont{L\"{o}wen}},\ }%
  \bibfield{journal}{%
  \Doi{10.1103/PhysRevLett.108.168301}{\bibinfo {journal} {Physical Review
  Letters}}\ }%
  \textbf{\bibinfo {volume} {108}},\ \bibinfo {pages} {168301} (\bibinfo
  {month} {Apr.}\ \bibinfo {year} {2012})%
  \bibAnnoteFile{NoStop}{bialke_crystallization_2012}%
\bibitem{Peruani2006}%
  \BibitemOpen
  \bibfield{author}{%
  \bibinfo {author} {\bibfnamefont{F.}~\bibnamefont{Peruani}}, \bibinfo
  {author} {\bibfnamefont{A.}~\bibnamefont{Deutsch}},\ and\ \bibinfo {author}
  {\bibfnamefont{M.}~\bibnamefont{B\"ar}},\ }%
  \bibfield{journal}{%
  \Doi{10.1103/PhysRevE.74.030904}{\bibinfo {journal} {Phys. Rev. E}}\ }%
  \textbf{\bibinfo {volume} {74}},\ \bibinfo {pages} {030904} (\bibinfo {month}
  {Sep.}\ \bibinfo {year} {2006})%
  \bibAnnoteFile{NoStop}{Peruani2006}%
\bibitem{Peruani2013}%
  \BibitemOpen
  \bibfield{author}{%
  \bibinfo {author} {\bibfnamefont{F.}~\bibnamefont{Peruani}}\ and\ \bibinfo
  {author} {\bibfnamefont{M.}~\bibnamefont{B\"ar}},\ }%
  \bibfield{journal}{%
  \bibinfo {journal} {New Journal of Physics}\ }%
  \textbf{\bibinfo {volume} {15}},\ \bibinfo {pages} {065009} (\bibinfo {year}
  {2013}),\ ISSN \bibinfo {issn} {1367-2630},\
  \url{http://stacks.iop.org/1367-2630/15/i=6/a=065009}%
  \bibAnnoteFile{NoStop}{Peruani2013}%
\bibitem{O'Hern2003}%
  \BibitemOpen
  \bibfield{author}{%
  \bibinfo {author} {\bibfnamefont{C.~S.}\ \bibnamefont{O'Hern}}, \bibinfo
  {author} {\bibfnamefont{L.~E.}\ \bibnamefont{Silbert}}, \bibinfo {author}
  {\bibfnamefont{A.~J.}\ \bibnamefont{Liu}},\ and\ \bibinfo {author}
  {\bibfnamefont{S.~R.}\ \bibnamefont{Nagel}},\ }%
  \bibfield{journal}{%
  \bibinfo {journal} {Phys. Rev. E}\ }%
  \textbf{\bibinfo {volume} {68}},\ \bibinfo {pages} {011306} (\bibinfo {year}
  {2003})%
  \bibAnnoteFile{NoStop}{O'Hern2003}%
\bibitem{Olsson2007}%
  \BibitemOpen
  \bibfield{author}{%
  \bibinfo {author} {\bibfnamefont{P.}~\bibnamefont{Olsson}}\ and\ \bibinfo
  {author} {\bibfnamefont{S.}~\bibnamefont{Teitel}},\ }%
  \bibfield{journal}{%
  \Doi{10.1103/PhysRevLett.99.178001}{\bibinfo {journal} {Phys. Rev. Lett.}}\
  }%
  \textbf{\bibinfo {volume} {99}},\ \bibinfo {pages} {178001} (\bibinfo {month}
  {Oct}\ \bibinfo {year} {2007}),\
  \url{http://link.aps.org/doi/10.1103/PhysRevLett.99.178001}%
  \bibAnnoteFile{NoStop}{Olsson2007}%
\bibitem{Ikeda2012}%
  \BibitemOpen
  \bibfield{author}{%
  \bibinfo {author} {\bibfnamefont{A.}~\bibnamefont{Ikeda}}, \bibinfo {author}
  {\bibfnamefont{L.}~\bibnamefont{Berthier}},\ and\ \bibinfo {author}
  {\bibfnamefont{P.}~\bibnamefont{Sollich}},\ }%
  \bibfield{journal}{%
  \Doi{10.1103/PhysRevLett.109.018301}{\bibinfo {journal} {Phys. Rev. Lett.}}\
  }%
  \textbf{\bibinfo {volume} {109}},\ \bibinfo {pages} {018301} (\bibinfo
  {month} {Jul}\ \bibinfo {year} {2012}),\
  \url{http://link.aps.org/doi/10.1103/PhysRevLett.109.018301}%
  \bibAnnoteFile{NoStop}{Ikeda2012}%
\bibitem{Henkes2011}%
  \BibitemOpen
  \bibfield{author}{%
  \bibinfo {author} {\bibfnamefont{S.}~\bibnamefont{Henkes}}, \bibinfo {author}
  {\bibfnamefont{Y.}~\bibnamefont{Fily}},\ and\ \bibinfo {author}
  {\bibfnamefont{M.~C.}\ \bibnamefont{Marchetti}},\ }%
  \bibfield{journal}{%
  \Doi{10.1103/PhysRevE.84.040301}{\bibinfo {journal} {Phys. Rev. E}}\ }%
  \textbf{\bibinfo {volume} {84}},\ \bibinfo {pages} {040301} (\bibinfo {month}
  {Oct.}\ \bibinfo {year} {2011})%
  \bibAnnoteFile{NoStop}{Henkes2011}%
\bibitem{Berthier2013b}%
  \BibitemOpen
  \bibfield{author}{%
  \bibinfo {author} {\bibfnamefont{L.}~\bibnamefont{Berthier}},\ }%
  \enquote{\bibinfo {title} {Nonequilibrium glassy dynamics of self-propelled
  hard disks},}\  (\bibinfo {month} {Jul.}\ \bibinfo {year} {2013}),\
  \url{http://arxiv.org/abs/1307.0704}%
  \bibAnnoteFile{NoStop}{Berthier2013b}%
\bibitem{Ni2013}%
  \BibitemOpen
  \bibfield{author}{%
  \bibinfo {author} {\bibfnamefont{R.}~\bibnamefont{Ni}}, \bibinfo {author}
  {\bibfnamefont{M.~A.~C.}\ \bibnamefont{Stuart}},\ and\ \bibinfo {author}
  {\bibfnamefont{M.}~\bibnamefont{Dijkstra}},\ }%
  \enquote{\bibinfo {title} {Pushing the glass transition towards random close
  packing using self-propelled hard spheres},}\  (\bibinfo {month} {Jun.}\
  \bibinfo {year} {2013}),\ \url{http://arxiv.org/abs/1306.3605}%
  \bibAnnoteFile{NoStop}{Ni2013}%
\bibitem{Fielding2013}%
  \BibitemOpen
  \bibfield{author}{%
  \bibinfo {author} {\bibfnamefont{S.}~\bibnamefont{Fielding}},\ }%
  \enquote{\bibinfo {title} {Hydrodynamic suppression of phase separation in
  active suspensions},}\  (\bibinfo {month} {Jul.}\ \bibinfo {year} {2013}),\
  \url{http://arxiv.org/abs/1210.5464}%
  \bibAnnoteFile{NoStop}{Fielding2013}%
\bibitem{Note1}%
  \BibitemOpen
  \bibinfo {note} {We have excluded runs at $\protect \mathaccentV
  {tilde}07E{\nu }_r=0$ in the glassy region from the analysis}%
  \bibAnnoteFile{NoStop}{Note1}%
\bibitem{Note2}%
  \BibitemOpen
  \bibinfo {note} {Since momentum is not conserved, the center of mass is not
  fixed. It moves with velocity ${\protect \bf v}_{cm}=\protect \mathrm
  {v}_0\DOTSB \sum@ \slimits@ _i {\protect \bm {n}}_i$ and $v_{cm}\sim \protect
  \mathrm {v}_0 N^{1/2}$. In a finite-size frozen or almost frozen state, this
  uniform drifting motion becomes the dominant contribution to the MSD and
  needs to be subtracted out.}%
  \bibAnnoteFile{Stop}{Note2}%
\bibitem{Donev2005}%
  \BibitemOpen
  \bibfield{author}{%
  \bibinfo {author} {\bibfnamefont{A.}~\bibnamefont{Donev}}, \bibinfo {author}
  {\bibfnamefont{F.~H.}\ \bibnamefont{Stillinger}},\ and\ \bibinfo {author}
  {\bibfnamefont{S.}~\bibnamefont{Torquato}},\ }%
  \bibfield{journal}{%
  \Doi{10.1103/PhysRevLett.95.090604}{\bibinfo {journal} {Phys. Rev. Lett.}}\
  }%
  \textbf{\bibinfo {volume} {95}},\ \bibinfo {pages} {090604} (\bibinfo {month}
  {Aug.}\ \bibinfo {year} {2005}),\
  \url{http://link.aps.org/doi/10.1103/PhysRevLett.95.090604}%
  \bibAnnoteFile{NoStop}{Donev2005}%
\bibitem{Note3}%
  \BibitemOpen
  \bibinfo {note} {Assuming particles' orientations are random and
  uncorrelated, the speed of a cluster of size $n$ is $V_n\sim n^{-1/2}$, and
  the relative speed of two clusters of sizes $n_1$, $n_2$ is $\Delta V\sim
  (n_1^{-1}+n_2^{-1})^{1/2}$}%
  \bibAnnoteFile{NoStop}{Note3}%
\bibitem{Note4}%
  \BibitemOpen
  \bibinfo {note} {Stiffness here refers to the force-distance repulsive
  scaling $F/k\sim r_{ij}^{-x}$ at small $r_{ij}$. Our harmonic potential
  remains finite, while e.g. $x=12$ for the WCA potential used by Redner et
  al.~\cite {Redner2013}.}%
  \bibAnnoteFile{Stop}{Note4}%
\bibitem{Note5}%
  \BibitemOpen
  \bibinfo {note} {Redner et al. define the P\'{e}clet number as $\protect
  \text {Pe}\protect \xspace _R=3\protect \text {Pe}\protect \xspace $. That
  puts our putative critical point at $\protect \text {Pe}\protect \xspace
  _R^*\simeq 30$, while they observe $\protect \text {Pe}\protect \xspace
  _R^*\simeq 50$ with WCA interactions and non-zero translational noise\cite
  {Redner2013}.}%
  \bibAnnoteFile{Stop}{Note5}%
\bibitem{Bialke2013}%
  \BibitemOpen
  \bibfield{author}{%
  \bibinfo {author} {\bibfnamefont{J.}~\bibnamefont{Bialk\'e}}, \bibinfo
  {author} {\bibfnamefont{H.}~\bibnamefont{L\"{o}wen}},\ and\ \bibinfo {author}
  {\bibfnamefont{T.}~\bibnamefont{Speck}},\ }%
  \enquote{\bibinfo {title} {Microscopic theory for the phase separation of
  self-propelled repulsive disks},}\  (\bibinfo {year} {2013}),\
  \url{http://arxiv.org/abs/1307.4908}%
  \bibAnnoteFile{NoStop}{Bialke2013}%
\bibitem{Tailleur2008}%
  \BibitemOpen
  \bibfield{author}{%
  \bibinfo {author} {\bibfnamefont{J.}~\bibnamefont{Tailleur}}\ and\ \bibinfo
  {author} {\bibfnamefont{M.~E.}\ \bibnamefont{Cates}},\ }%
  \bibfield{journal}{%
  \Doi{10.1103/PhysRevLett.100.218103}{\bibinfo {journal} {Phys. Rev. Lett.}}\
  }%
  \textbf{\bibinfo {volume} {100}},\ \bibinfo {pages} {218103} (\bibinfo
  {month} {May}\ \bibinfo {year} {2008})%
  \bibAnnoteFile{NoStop}{Tailleur2008}%
\bibitem{reichhardt_dynamical_2011}%
  \BibitemOpen
  \bibfield{author}{%
  \bibinfo {author} {\bibfnamefont{C.}~\bibnamefont{Reichhardt}}\ and\ \bibinfo
  {author} {\bibfnamefont{C.~J.~O.}\ \bibnamefont{Reichhardt}},\ }%
  \bibfield{journal}{%
  \Doi{10.1073/pnas.1116359108}{\bibinfo {journal} {Proceedings of the National
  Academy of Sciences}}\ }%
  \textbf{\bibinfo {volume} {108}},\ \bibinfo {pages} {19099 } (\bibinfo
  {month} {Nov.}\ \bibinfo {year} {2011}),\
  \url{http://www.pnas.org/content/108/48/19099.short}%
  \bibAnnoteFile{NoStop}{reichhardt_dynamical_2011}%
\bibitem{Berthier2013a}%
  \BibitemOpen
  \bibfield{author}{%
  \bibinfo {author} {\bibfnamefont{L.}~\bibnamefont{Berthier}}\ and\ \bibinfo
  {author} {\bibfnamefont{J.}~\bibnamefont{Kurchan}},\ }%
  \bibfield{journal}{%
  \Doi{10.1038/nphys2592}{\bibinfo {journal} {Nat Phys}}\ }%
  \textbf{\bibinfo {volume} {9}},\ \bibinfo {pages} {310} (\bibinfo {month}
  {May}\ \bibinfo {year} {2013}),\ ISSN \bibinfo {issn} {1745-2473},\
  \url{http://www.nature.com/nphys/journal/v9/n5/full/nphys2592.html}%
  \bibAnnoteFile{NoStop}{Berthier2013a}%
\bibitem{Note6}%
  \BibitemOpen
  \bibinfo {note} {This is also the basis for the interface stability argument
  in Ref.~\protect \bibpunct {}{}{,}{n}{}{}\cite {Redner2013}\protect \bibpunct
  {}{}{,}{s}{}{}.}%
  \bibAnnoteFile{Stop}{Note6}%
\end{thebibliography}%

\end{document}